\documentclass[sigconf]{acmart}

\AtBeginDocument{%
  }

\copyrightyear{2026}
\acmYear{2026}
\setcopyright{cc}
\setcctype{by}
\acmConference[CHI '26]{Proceedings of the 2026 CHI Conference on Human Factors in Computing Systems}{April 13--17, 2026}{Barcelona, Spain}
\acmBooktitle{Proceedings of the 2026 CHI Conference on Human Factors in Computing Systems (CHI '26), April 13--17, 2026, Barcelona, Spain}
\acmPrice{}
\acmDOI{10.1145/3772318.3791732}
\acmISBN{979-8-4007-2278-3/2026/04}

\usepackage{arabtex}
\usepackage{utf8}
\setcode{utf8}
\usepackage{array,booktabs,ragged2e}
\usepackage{subcaption}
\usepackage{placeins}
\newcolumntype{R}[1]{>{\RaggedLeft\arraybackslash}p{#1}}

\newcommand{\coloring}[1]{\textcolor{black}{#1}}

\newcommand{\change}[2]{\coloring{#1}\iffalse{#2}\fi}

\begin{document}

\title{``God says we are right!'': The Interplay between Religion and Propaganda on Arabic Social Media}

\author{Mahmoud Fawzi}
\email{M.F.G.Ibrahim@sms.ed.ac.uk}
\orcid{0009-0008-8239-3333}
\affiliation{%
  \institution{The University of Edinburgh}
  \city{Edinburgh}
  \country{UK}
}

\author{Björn Ross}
\email{b.ross@ed.ac.uk}
\orcid{0000-0003-2717-3705}
\affiliation{%
  \institution{The University of Edinburgh}
  \city{Edinburgh}
  \country{UK}
}

\author{Walid Magdy}
\email{wmagdy@inf.ed.ac.uk}
\orcid{0000-0001-9676-1338}
\affiliation{%
  \institution{The University of Edinburgh}
  \city{Edinburgh}
  \country{UK}
}

\begin{abstract}
Religion is a main aspect of life in many parts of the world; hence, it acts as a powerful tool for influencing people's views and actions. In the context of propaganda and misinformation, religion has been perceived as a factor that impacts people negatively leading them to be influenced by false or biased information. However, there are limited quantitative studies that explore the exact role of religion in this process. In this study, we investigate whether state-sponsored propaganda accounts share religious content in a different way than the typical norms reported in the literature and whether they mobilize this content to promote their agendas. This is achieved by exploring 15 Middle Eastern propaganda Twitter datasets encompassing around 124 million tweets from 32.5K accounts attributed to KSA, UAE, Egypt, Turkey, and Iran. We find that propagandists share excerpts from the Quran and Hadith according to patterns that reflect aspects of their state's agenda. We also demonstrate examples where such religious texts were used to modulate political messages.
\end{abstract}



\maketitle

\section{Introduction}
The involvement of religion in the systematic manipulation of public opinion, known now as propaganda, has existed since the ages of ancient empires \cite{wilkinson2002early, portefaix1993ancient, diamond2003propaganda}. Historically, religious propaganda was an important aspect in deadly conflicts such as The French Wars of Religion \cite{racaut2017hatred}, The Thirty Years War \cite{gilly2000midnight}, and Taiping Rebellion \cite{boardman1951christian} which left millions of casualties behind. Lately, a set of recent events increased  public and the academic interest in the religious dimensions of political conflict \cite{brubaker2015religious}. For example, the Russian absorption of Crimea was linked multiple times by  Russian propaganda to religious roots such as the baptism of prince Vladimir there \cite{kozelsky2014religion}. The effect of the connection between religion and propaganda goes beyond political dynamics and agendas, also affecting the religious practice of individuals. \citet{salnikova2025role} report many religious transfers from the Ukrainian Orthodox Church of the Moscow Patriarchate (UPC MP) to the Orthodox Church of Ukraine (OCU) amid the Russo-Ukrainian war. \change{In the context of HCI research, multiple social studies report that sharing religious content on social media is imposing new challenges to religious discussions online, which urges social media platforms and users to such content in a special way \cite{kaplan2023religious, simamora2024rethinking, bowman2025sharing, Biswas2025SocialMediaReels}. However, few HCI studies address these concerns \cite{rifat2022putting, rifat2022situating}. More importantly, there is very limited work that investigates how sharing religious content on social media contributes to aspects like undermining democracy and intensifying conflicts \cite{fawzi2026holiness}.}{change-R2C3}

The Middle East is one of the most religious regions in the world \cite{evans}. In addition, it encompasses four out of the most ten dangerous countries in the world namely, Palestine, Syria, Lebanon, and Sudan \cite{acled}. Over the past decade, the region has witnessed the "Arab Spring", which was a series of uprisings and armed rebellions in which political Islam had a significant role \cite{esposito, ferrero}. Islam is the dominant religion in the MENA region, with 94\% of the population being Muslims \cite{hackett}. These factors motivated our choice for studying the potential interplay between Islam and propaganda in this highly understudied region.

\begin{figure*}
\centering
\begin{subfigure}{.33\textwidth}
  \centering
  \includegraphics[width=0.8\linewidth]{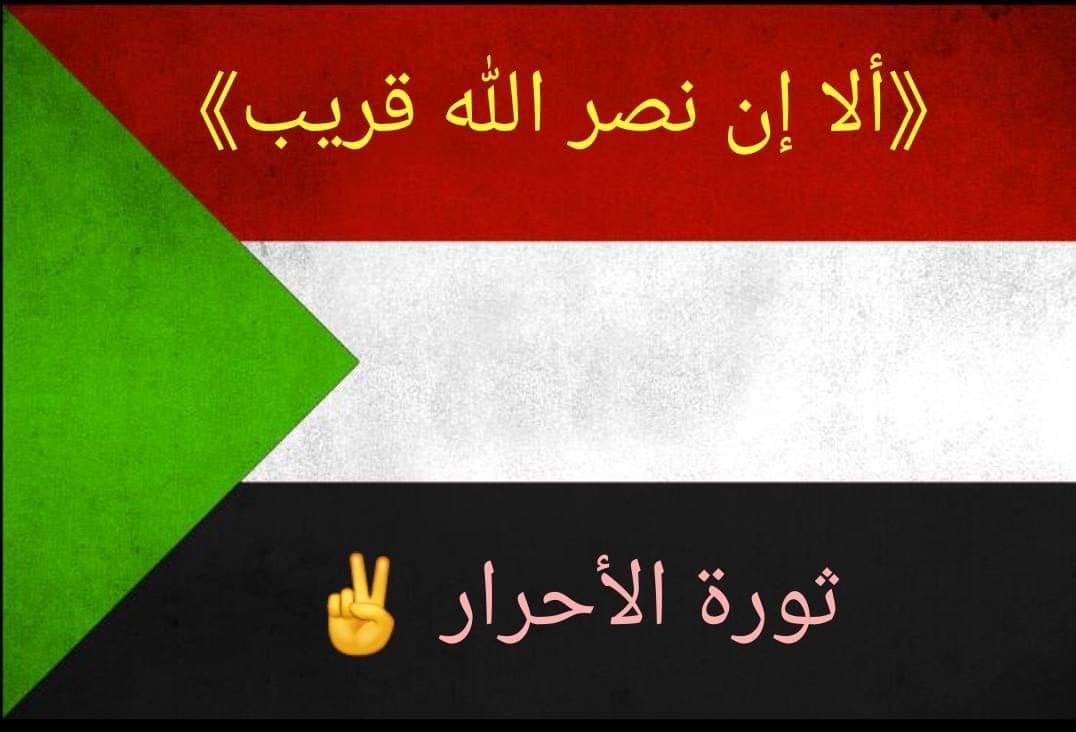}
  \caption{}
  \label{fig:example1}
\end{subfigure}%
\begin{subfigure}{.33\textwidth}
  \centering
  \includegraphics[width=0.8\linewidth]{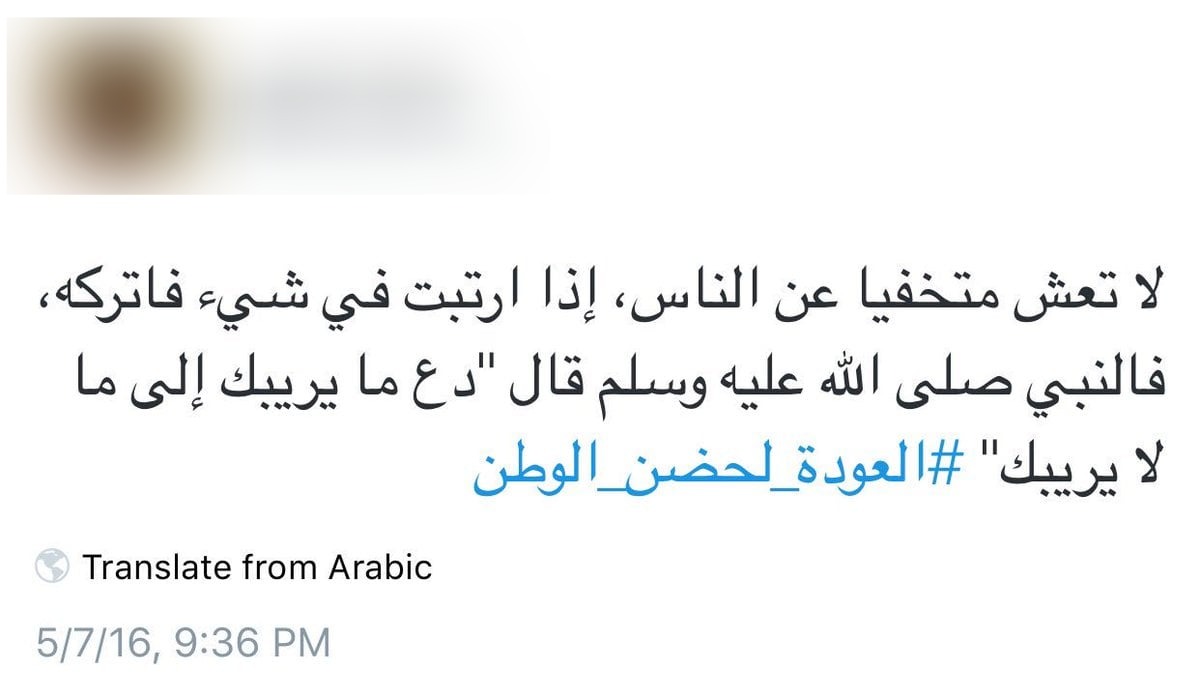}
  \caption{}
  \label{fig:example2}
\end{subfigure}
\begin{subfigure}{.33\textwidth}
  \centering
  \includegraphics[width=0.7\linewidth]{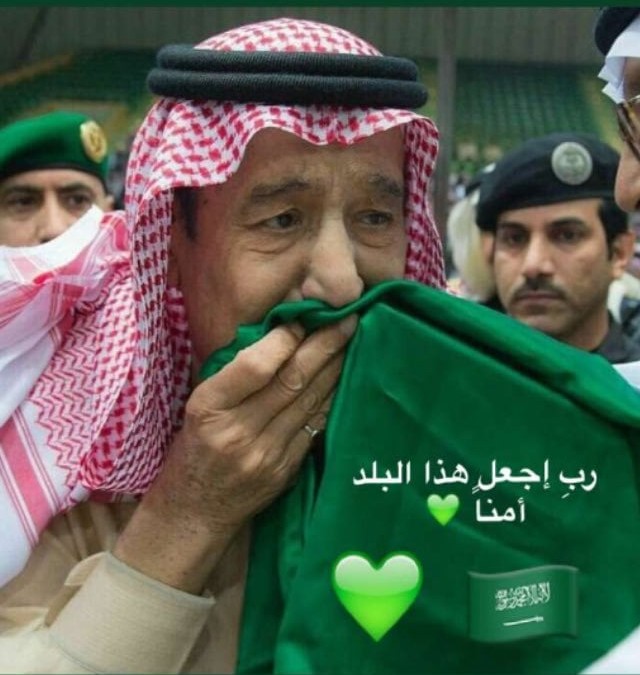}
    \caption{}
  \label{fig:example3}
\end{subfigure}
\caption{Examples of using Quran and Hadith for supporting political messages: \\\textbf{(a)} Sudan's flag with the Quran verse \textit{Indeed, God’s victory is near} and the slogan \textit{Free people's revolution}. \\ \textbf{(b)} The hadith \textit{Give up what is doubtful for what is not doubtful} and a hashtag supporting UAE's government. \\ \textbf{(c)} A photo of KSA's king kissing its flag with the Quran verse \textit{My Lord! Make this country safe}.}
\label{fig:examples}
\Description{Examples of using Quran and Hadith for supporting political messages: \\\textbf{(a)} Sudan's flag with the Quran verse \textit{Indeed, God’s victory is near} and the slogan \textit{Free people's revolution}. \\ \textbf{(b)} The hadith \textit{Give up what is doubtful for what is not doubtful} and a hashtag supporting UAE's government. \\ \textbf{(c)} A photo of KSA's king kissing its flag with the Quran verse \textit{My Lord! Make this country safe}.}
\end{figure*}

Islam is followed by 24.9\% of the world population, making it the second most followed religion worldwide \cite{cia}. It relies on two major sources: the Quran (its holy book), and Hadith (the record of the words and actions of the prophet Muhammad) \cite{binbeshr2021systematic}. The use of both Quran \cite{clausen2023unity} and Hadith \cite{glausch2020infographics, boutz2019exploiting} for propaganda is well-documented, for example by the Islamic State of Iraq and Syria (ISIS) \change{to assert legitimacy}{change-R2C3}. However, the scope of these studies is limited to analyzing a very small portion of such incidences as they focus on the official releases and the main channels of ISIS only without checking further how those messages propagated on social networks and among the target audience. Additionally, they do not investigate how other players in the region could have potentially used Quran and Hadith to support their propaganda and serve their own agendas. This is particularly important because multiple studies about the Middle East show that the ecosystem of propaganda and misinformation in the region relies on state-sponsored accounts \cite{voiceFromAbove, fawzi2024pinocchio} \change{and that religion has a role in this process as people highly respect it \cite{rolston2020everywhere}. Furthermore, some other studies report a different role of religion in the Middle East where religious welfare organizations maintain social order through social services \cite{jawad2025critical}.}{change-R2C3}

Figure \ref{fig:examples} shows three examples from Twitter's released propaganda datasets \cite{twitter_transparency} where accounts linked with the states of Egypt and UAE use religious texts to support political messages. In Figure \ref{fig:example1}, a Quran verse is used to support the Sudanese revolution that started in 2018. Figure \ref{fig:example2} shows a tweet quoting a hadith along with the hashtag \<العودة لحضن الوطن>\# \textit{(\#back\_to\_homeland)} which is linked with a series of posts asking people to refrain from supporting political Islam and to support the state's government instead. Figure \ref{fig:example3} shows the King of Saudi Arabia kissing its flag along with a Quran verse. We try through our study to dig deeper into this phenomenon and answer the following research questions:
\begin{itemize}
    \item \textbf{RQ1:} What religious texts, if any, do state-sponsored propaganda accounts share and how does this differ from what is reported in the literature about how religious texts are shared on social media in general?
    \item \textbf{RQ2:} Do these state-sponsored propaganda accounts use religious texts to promote their agenda, and if so, how?
\end{itemize}

We use a quantitative approach to answer \textbf{RQ1} and a \change{mixed-methods}{} approach \change{(quantitative \& qualitative)}{}  to answer \textbf{RQ2}. We address these points by processing 124 million tweets \change{from 32.5K accounts}{change-WRequest} originating from the state-sponsored operations of five influential countries in the Middle East: Iran, Saudi Arabia, Egypt, United Arab Emirates, and Turkey. These tweets were released by Twitter (now called X) as a part of its Twitter Transparency Report project \cite{twitter_transparency}. First, we use established methods from the literature to identify Arabic tweets that quote texts from the Quran or the Hadith \cite{abokhodair2020holy, fawzi2025prophet}. Then, we check how far the patterns of sharing these religious texts by state-sponsored propaganda accounts align with the patterns reported in the literature. At last, we analyze more closely the instances where those accounts used religious texts to spread political messages that support their agendas.

Our findings confirm that state-sponsored propaganda accounts in the Middle East share Quran and Hadith regularly. Perhaps surprisingly, this is not only the case for Iran, which is widely described as a theocracy \cite{parchami2024neither, ghobadzadeh2016electoral}. It is also the case for Saudi Arabia (KSA), the United Arab Emirates (UAE), and Egypt whose main \textit{raison d'état} is to counter political Islam \cite{roll2023egypt}. In addition, we observe that the most shared religious texts by each country as well as its sharing pattern in terms of topical distribution conform with its state's agenda. Finally, for each of the countries we study, we spot within the state-sponsored tweets clear political messages that leverage texts from the Quran or the Hadith and provide interpretations that are intended to support their claims.

We believe that our results are very relevant to the HCI research community as they constitute the first computational outcome that quantifies religious propaganda on social media. In addition, our work is the first study to examine the religious dimension of the state-sponsored operations in the Middle East, an understudied, highly volatile part of the world. In the context of \change{building better social media platforms}{change-R1C1}, our results pinpoint an understudied form of social media misuse that features complex structures of manipulation. \change{This contributes to the limited work that demonstrates design gaps in social media platforms that distance them from handling the complexities of religious texts \cite{aahman2021facebook, rifat2022putting}.}{change-R1C1}

\section{Related Work}

In this section, we introduce some related work in HCI and social sciences that has tackled the digital presence and interactions involving religion. Then, we present some studies that analyzed state-sponsored accounts suspended by Twitter. At last, we show some works that emphasize the interplay between religion and content moderation on social media.

\subsection{Digital Religion}
Religious and spiritual practices and discussions are now increasingly mediated by technology \cite{wolf2024still, campbell2021digital, tsuria2017approaches}. This applies to Christianity \cite{hutchings2015reading, hutchings2017design, hutchings2017creating}, Judaism \cite{tsuria2018understanding, tsuria2016jewish, tsuria2017sin, tsuria2020discourse}, Islam \cite{fakhruroji2019digitalizing, abokhodair2020holy}, Buddhism \cite{claisse2023keeping}, and beyond. Since conversational AI systems have become the most trending technologies during the recent years, thoughts about how they would reshape religious practice started to arise. A multi-modal embodiment of conversational AI inspired by the beliefs and symbols of Korean shamanism was introduced and tested in a recent study that reported positive feedback provided by participants \cite{cho2025shamain}. \citet{alam2025blind} conducted a study to assess AI-generated religious content and revealed a significant discrepancy. While Muslims expressed distrust in AI for religious guidance and stated a preference for scholarly answers, they preferred AI-generated responses to Islamic questions in blind evaluations. This does not mean that such systems are ready to serve as religious references for Muslims as the same study revealed critical deficiencies in AI responses, highlighting users’ limited ability to recognize such shortcomings. \citet{mubarak2025islamiceval} have recently created a shared task with the aim of capturing the hallucinations of Large Language Models (LLMs) in Islamic content.

Besides LLMs, researchers are still attempting to enhance the religious practice experience through physical interactive technologies \cite{song2025walking, claisse2023keeping, kim2022social, markum2023designing}. An example involving Muslim women was the work of \citet{ibrahim2024tracking} that showed the importance of designing personal informatics tools tailored for them to track their health during the special Islamic month Ramadan. This work was extended later to propose more tailored features that can also be useful apart of Ramadan, covering menstruation and pregnancy healthcare \cite{ibrahim2024expanding, ibrahim2024islamically}.

Nevertheless, literature reviews covering HCI research on digital religion \cite{buie2013spirituality, wolf2024still} still consider the amount of research on the topic very little. More importantly, they pinpoint an even further lack in studies that contribute social knowledge to the community rather than proposing experimental designs \cite{wolf2024still}. An example of these studies \change{that show the potential for weaponizing religion}{change-R2C4} is the work of \citet{al2021atheists} that studies religious polarization in Arab online communities and examines how \change{religious perceptions cause social divide as well as clusters of people ending up in echo chambers}{change-R2C4}. \change{Another relevant example is the work of \citet{borge2015content} that tackles political polarization in Egypt and how Islamism served as a political ideology that shapes opinion evolution dynamics.}{change-R2C4} A recent study examined the interactions of Arab users with fabricated Hadith \cite{fawzi2026holiness}. They found that the phenomenon of the spread of fabricated Hadith is manifesting a lot of observations about the different cultural backgrounds of social media users in the Middle East and the disputes they have in a religious flavor. They found instances of fabricated Hadith  that are deployed in the Sunni-Shia sectarian conflict \cite{mcmahon, alshehri2023comfort}. The lack of this kind of studies, especially those that address an understudied language (Arabic) within an understudied community (Muslims) in an understudied region (Middle East), was a main motive behind our study. \change{These sample studies, among others, naturally pose questions about the exact role of religion in manipulating people and whether this influence is planned and coordinated or not.}{change-R2C4}

\subsection{State-sponsored Accounts}
Twitter (currently X) used to periodically share sets of accounts that have been suspended because they were found to be linked to state-sponsored activities for the purpose of promoting Propaganda. These sets of accounts were shared under a project called Twitter Transparency Report which was discontinued in 2022 \cite{twitter_transparency}. The research community has studied the activity of these accounts in multiple ways. \citet{guo2022large} augmented the dataset by providing a negative set of accounts that are \textbf{not} state-backed from Twitter Archive to allow researchers to develop systems that can distinguish them. \citet{luceri2024unmasking} utilize network properties to identify coordinated influence operations (IO) drivers. They apply network pruning and generate a vector representation of the fused similarity network before using a supervised learning model for detection. Nevertheless, the study picks only one dataset for Arab countries that has accounts from Egypt and UAE together. These accounts from Egypt and UAE were investigated again by \citet{wang2023evidence}, who had a similar focus on coordination. The main objective of the study was to prove and analyze the coordination among state-backed information operations. The study however does not dig deeper into the nature of the content shared by these accounts beyond reporting the top hashtags found. Generally, this set of studies approach those accounts through a network analysis lens and not through a social lens that engages with the challenges of the HCI community.

A deeper investigation of the content shared by the Russian state-sponsored accounts was carried out by \citet{zannettou2019disinformation}. Their behavior is compared against a random set of Twitter users in terms of the temporal characteristics, top words shared, client devices used, and other features like top hashtags, top mentions, and top domains. After detecting a significant use of trolls within the activity of these accounts, this work was extended to analyze those images and trolls further \cite{zannettou2020characterizing, mejova2025narratives}. Our study has a similar nature to these studies; however, instead of examining trolls, we focus on the religious component within the content of these accounts. Instead of analyzing Russian accounts, we focus on the big portion of understudied accounts created by Middle Eastern countries to manipulate people in this volatile yet very understudied part of the world. \change{The impact of religious texts in this region has already been acknowledged by earlier qualitative studies that showed how ISIS used them to assert its legitimacy \cite{clausen2023unity, boutz2019exploiting}, however, such studies did not consider other actors apart from ISIS nor analyzed this weaponization at scale.}{change-R2C4} The limited work dedicated to this region on those datasets avoids addressing the sociopolitical aspects of propaganda and performs analysis to gain insights about other topics like sports and banking \cite{almotairy2024characterizing}.

\subsection{Religion \& Content Moderation}
As religion influences the dynamics of social media and the discussions on it \cite{boutz2017quoting}, the inherent issues of social media like hate speech \cite{castano, albadi2019hateful, albadi, ali2022analyzing}, misinformation \cite{alimardani2020covid19, lee2022covid}, and polarization \cite{al2021atheists} also influence religious groups. Deradicalization is often proposed as a countermeasure against online religious hate speech \cite{albadi2022deradicalizing, kursuncu2019modeling}. The process involves a pipeline that relies on modeling, characterization, detection, and personalization. \citet{chandra2021virus} created a dataset called \textit{CoronaBias} that includes incidences of Islamophobic behavior triggered by the COVID-19 crisis. \citet{al2024social} tackled the issue of religious misinformation and highlighted the topical categories in which people get deceived and manipulated the most. Similarly, \citet{fawzi2026holiness} demonstrated the phenomenon of fabricating religious texts, particularly Hadith. They showed clear differences between the networks of accounts circulating fabricated religious texts and those trying to debunk this misinformation.

We believe in the continuous need of similar studies that revisit content moderation strategies and challenges to cover the gap created by the rapid change and development in technology. We contribute to the understanding of the spread of religious propaganda online to assist creating moderation techniques that can counter it.

\section{Data \& Methods}

We start this section by introducing the details of the datasets used in this work. Then, we describe how religious texts are extracted and analyzed further.

\subsection{Data Description}

Table \ref{tab:dataset-statistics} shows the countries involved in our study along with the number of datasets, accounts, and tweets. Initially, we included all the Arab countries spotted in the Twitter Transparency Report datasets namely, Saudi Arabia (KSA), United Arab Emirates (UAE), and Egypt. We also included two datasets which contained accounts from multiple coordinating countries (a dataset involving Egypt and UAE and a separate dataset involving KSA, UAE, and Egypt). Finally, we considered Turkey and Iran in our analysis. Although both countries do not belong to the Arab league, both of them are Middle Eastern countries that have a Muslim majority and a record of political intervention in the politics of the Arab region.

\begin{table}
\begin{center}
\resizebox{0.48\textwidth}{!}{
\begin{tabular}{l c c c c c c} 
 \hline
  & KSA & UAE & Egypt & Turkey & Iran & Coordination \\
 \hline
 Datasets & 3 & 1 & 1 & 1 & 7 & 2 \\ 
 Accounts & 5,968 & 4,248 & 2,541 & 7,340 & 6,787 & 5,621 \\
  Tweets & 32.3M & 1.3M & 7.9M & 36.9M  & 9.1M  & 36.5M \\ 
 \hline
\end{tabular}
}
\end{center}
    \caption{The number of datasets, accounts, and tweets per country in our data collection.}
    \label{tab:dataset-statistics}
\end{table}

A country might have multiple datasets because each of them comprises a network of accounts that have been suspended by Twitter at a point of time after their suspicious activity was detected and validated. For most datasets, Twitter wrote blog posts justifying why the activity of these accounts was regarded as violating the community standards. We summarize these blogs below:

\begin{itemize}
    \item \textbf{KSA:} In September 2019, Twitter announced the suspension of six accounts linked to Saudi Arabia’s state-run media apparatus because they presented themselves as independent journalistic outlets while tweeting narratives favorable to the Saudi government. Later in December 2019, a group of 5,929 accounts were also suspended because they represented the core portion of a larger network of more than 88,000 accounts engaged in spammy behavior across a wide range of topics. They amplified Saudi propaganda through inauthentic engagement tactics such as aggressive liking, retweeting and replying. While the majority of the content from this network was in Arabic, a portion of it related to events relevant to Western audiences, including amplification of discussion around sanctions in Iran and appearances by Saudi government officials in Western media. As per Twitter, many of the accounts employed automated tools to also share non-political content at high volumes to mask the overall manipulation they are doing. Twitter's investigations traced the source of this network to Smaat\footnote{Smaat Company: \url{linkedin.com/company/smaat-co}} (in Arabic \<سماءات>), a social media marketing and management company based in Saudi Arabia. Finally, in October 2020, another 33 accounts originating from Saudi Arabia with ties to the Saudi government were suspended because they were impersonating key Qatari political figures to advance narratives about Qatari politics which are geostrategically favorable to the Saudi authorities.
    \item \textbf{UAE:} In the same blog of September 2019, the suspension of 4,248 operating from the UAE was reported. The accounts were mainly directed at Qatar and Yemen, employing false personae and tweeting  about regional issues, such as the Yemeni Civil War and the Houthi Movement.
    \item \textbf{Egypt:} In April 2020, an Egyptian network of 2,541 accounts linked to the pro-government El Fagr newspaper\footnote{El Fagr Newspaper: \url{elfagr.org}} (in Arabic \<الفجر>), was also removed. Twitter said that the fake accounts had been used to amplify messaging critical of Iran, Qatar and Turkey.
    \item \textbf{Turkey:} In June 2020, Twitter disclosed 7,340 fake and compromised accounts which were being used to amplify political narratives favorable to the AK Parti, and demonstrated strong support for President Erdogan. Technical signals associated the network with the youth wing of the AK party and a centralized network that maintained a significant number of compromised accounts. These compromised accounts were associated with organizations critical of President Erdogan and the Turkish Government; hence, they have been repeated targets of account hacking and takeover efforts by the state actors identified above. The broader network was also used for commercial activities, such as cryptocurrency-related spam. 
    \item \textbf{Iran:} Iranian networks were regular guests in the transparency reports of Twitter. In October 2018, 770 accounts originating from Iran were disclosed due to an attempted influence campaign. By January 2019, another 2,617 accounts were suspended for the same reason. In June 2019, Twitter announced the suspension of 3 different Iranian networks. The first one had 1,666 accounts who shared global news with an angle that benefited the diplomatic and geostrategic views of the Iranian state. The second set had 248 accounts engaged with discussions related to Israel specifically. The third set comprised 2,865 accounts and employed a range of false personas to target conversations about political and social issues in Iran and globally. In October 2020's blog, the suspension of 104 Iranian accounts was reported as they artificially amplified conversations on politically sensitive topics, including Black Lives Matter (BLM), the murder of George Floyd, and other issues of racial and social justice in the United States. Finally, in February 2021, 238 additional accounts were suspended. It is worth noting that the sum of these counts (8,508) does not match exactly the number of accounts in the released datasets (6,787) due to some redundancies in the released files.
    \item \textbf{Coordination Datasets:} In September 2019, 271 accounts originating from  UAE and Egypt were removed after finding evidence that they had been created and managed by a private technology company operating in both countries called DotDev. In addition to targeting Qatar and Iran, the accounts interestingly amplified messages supportive to the Saudi government confirming the alliance reported by \citet{roll2023egypt}. Later in April 2020, 5,359 accounts operating from KSA, UAE, and Egypt were removed after praising the Saudi leadership, criticizing Qatar and Turkish activity in Yemen, criticizing the Syrian government, and supporting the Libyan warlord Khalifa Haftar. The activity of this dataset was analyzed further by the Stanford Internet Observatory\footnote{Stanford Internet Observatory Report: \url{cyber.fsi.stanford.edu/io/news/april-2020-twitter-takedown}}.
\end{itemize}

Despite the availability of these official blogs, there is a very little attention and analysis to this comprehensive data collection which reveals a lot of facts about the dynamics of alliances and rivalries in the Middle East.

\subsection{Extraction of Islamic Texts}
To explore the role of religious texts in these propaganda campaigns, we apply filtering to our tweets to keep only the tweets that include excerpts from the Quran and/or the Hadith. While the quotes of some historic Islamic characters other than the prophet are also valuable for many Muslims and can be used for manipulation, we stick to those two sources as they have a solid methodology for extracting them in previous works. We filter the tweets using the \textit{tweet\_language} field to consider Arabic tweets only. Then, we apply the same methods reported in earlier studies \cite{abokhodair2020holy, fawzi2025prophet} so that we can compare the sharing patterns produced by state-sponsored propaganda accounts to the patterns reported in the literature.

\begin{table}[]
    \centering
    \small
    \begin{tabular}{c c}
    \hline
         Key Phrase & Translation  \\
         \hline
         \<بسم الله الرحمن الرحيم> & In the name of God \\ \<قال تعالى> & God said \\
         \<صدق الله العظيم> & God Almighty has spoken the truth \\ \<قال المولى> & God said \\
         \<قوله تعالى> & God's saying \\ \<قال عز وجل> & God Almighty said \\
         \<قال في كتابه> & God said in his book \\
         \hline
    \end{tabular}
    \caption{The keywords used to extract excerpts from the Quran.}
    \label{tab:quran-keywords}
\end{table}

\subsubsection{Quran}
For extracting Quran from tweets, we follow the algorithm developed by \citet{abokhodair2020holy}. First we search all the tweets for any of the key phrases listed in table \ref{tab:quran-keywords}. Then, we split each tweet into parts using punctuation and newlines. For each part that has more than three words, we check if it is a substring of a Quran verse. It is possible for a tweet part to belong to multiple Quran verses. In fact, there are multiple Quran verses that are identical but exist in different positions in the book.

\subsubsection{Hadith}
For extracting Hadith, we follow the approach described by \citet{fawzi2025prophet}. We search all the tweets for the keyword \<قال رسول الله> which means "The Messenger of God said". Afterwards, we search in these tweets for hadiths provided by two Hadith datasets namely, MAHADDAT \cite{gaanoun} and the LK Corpus \cite{altammami}. The search however does not follow a basic substring matching algorithm because of the flexible nature of Hadith as a language resource \cite{fawzi2025prophet}. Instead, both datasets as well as the tweets undergo a matching pipeline that includes multiple steps of preprocessing, hashing, and Jaccard similarity calculation. Then, each tweet is considered to include the hadith with the highest computed Jaccard similarity for it if this Jaccard similarity score is above an experimented threshold (0.35).

\subsubsection{\coloring{Data Validation}}

\change{We manually investigated the precision of the identified Quran and Hadith tweets by randomly investigating
100 tweets of each. All the tweets in the potential Hadith sample included hadiths showing a precision of 100\% (P=1). For the Quran sample, only 95 tweets included Quran (P=0.95), while four tweets included the phrase "In the name of God" which is a Quran verse but is also used in different contexts by Muslims; and one tweet, which included the phrase "God said", was followed by a special type of Islamic text called \texttt{Hadith Qudsi} \cite{zwemer1922so}. These are around 40-50 hadiths in Islam that are words of God that are said with Muhammad's wording.}{change-R1C3} 

\subsubsection{\coloring{Data Limitations}}
\change{While these algorithms for identifying Quran and Hadith in tweets showed very high precision, we acknowledge that they might have some limitations when it comes to recall. For example, the algorithm used by \citet{abokhodair2020holy} ignores matching Quran verses and fragments less than three words, which would ignore matching verses with one or two words only. Also, The Hadith extraction algorithm by \citet{fawzi2025prophet} is reported to achieve a recall of around 75\% to preserve the high precision. Nevertheless, we stick to both algorithms to mitigate the bias of changing the algorithms in the results and to make our findings more comparable to the literature.}{change-R1C3}

\change{It is also important to acknowledge that despite using the same language and the same algorithms of earlier studies, our results are not completely comparable to what is reported by \citet{abokhodair2020holy} and \citet{fawzi2025prophet} because their data are drawn from different distributions with a different date range, which limits us from generating a control dataset. Nonetheless, we still think that examining the differences in the sharing patterns between propagandists and normal accounts is providing valuable insights despite these limitations since the tweets on both sides are in the order of hundreds of thousands, covering multiple years, and covering all the months of these years. More importantly as we show later, the disproportions that we report are not just deviations from the sharing patterns of normal accounts but also from the original distributions of Quran and Hadith per topic apart of social media.}{change-R1C3, change-R3C2}

\change{\subsection{Thematic Analysis}}{change-R1C2}

\begin{table}[]
    \centering
    \small
    \begin{tabular}{c c | c c }
    \hline
         Keyword & Translation & Keyword & Translation \\
         \hline
         \<سعوديه>\ , \<سعودية> & Saudi Arabia & \<امارات>, \<إمارات> & UAE \\ \<مصر> & Egypt &
         \<تركيا>\ & Turkey \\ \<إيران>, \<ايران> & Iran & \<قطر>\ & Qatar \\
         \<سوريا> & Syria & \<ليبيا> & Libya \\ \<اليمن> & Yemen &
         \<لبنان> & Lebanon \\ \<سودان> & Sudan \\
         \<سلمان> & Salman & \<زايد> & Zayed \\ \<سيسي>, \<سيسى> & Sisi &
        \<اردوغان> & Erdogan \\ \<خامنئي> & Khamenei & \<تميم> & Tamim \\
        \<بشار> & Bashar & \<حفتر> & Haftar \\ \<حوثي> & Houthi &
        \<حزب الله> & Hezbollah \\ \<حسن نصر الله> & Hassan Nasrallah \\
         \hline
    \end{tabular}
    \caption{The keywords used to extract political tweets.}
    \label{tab:political-keywords}
\end{table}

\subsubsection{Political Tweets}
Table \ref{tab:political-keywords} shows the keywords used to extract the political tweets that we analyze further to answer our \textit{RQ2}. They include basically the names of the countries that have spread the propaganda, the names of the countries targeted by propaganda, and the names of political leaders of the countries in each set. Some of the words are written in two variants because Arabs use them interchangeably although one of them is considered a spelling mistake according to Modern Standard Arabic (MSA). We are aware that these tweets do not represent a comprehensive list of all the political tweets in the \change{sampling frame}{change-R2C2} as some political tweets might involve arguments and discussions that do not include these keywords. Nevertheless, they still provide a useful sample of the political messages.

\begin{table*}[]
\begin{center}
\begin{tabular}{p{1.2cm} p{14cm} p{1.5cm}} 
 \hline
  Theme & Description & Reference \\
 \hline
 Support & \emph{Supporting one's own country or its allies}, which includes the following tactics or a combination of them: 1) Quoting a religious text that explicitly praises a target country or group or that praises a historic or religious person or group then mapping this to an existing person or group. 2) Justifying political decisions by relating them to religious texts. 3) Attaching a support hashtag to a religious text.& 1, 3, 7, 9, 10, 15 \\ 
 Attack & \emph{Criticism of opponents}, which includes the following tactics or a combination of them: 1) Quoting a religious text that explicitly criticizes a target country or group or that criticizes a historic or religious person or group then mapping this to an existing person or group. 2) Interpreting past events and foreseeing future events based on interpretations of religious texts. 3) Attaching a criticism hashtag to a religious text.& 2, 4, 5, 6, 8, 11, 12, 13, 14, 16 \\
Non-political & Tweets of religious texts that include political keywords with a different meaning or tweets of religious texts that include hashtags of names of countries in a neutral way. & 17, 18 \\
 \hline
\end{tabular}
\end{center}
    \caption{\change{The themes extracted from the annotation of data and their descriptions. Reference examples can be found in table \ref{tab:political-tweets}.}{change-R1C2}}
    \label{tab:codebook}
\end{table*}

\subsubsection{\coloring{Codebook}}
\change{We follow the well-known reflexive thematic analysis process introduced by \citet{braun2006using} which they described later in more detail in their book \cite{braun2021thematic} so that we identify themes that are relevant to our research questions. Two authors of our work are native Arabic speakers and they performed the annotation process. One of them goes through \textbf{all} the extracted potential political tweets to perform the initial phase of familiarization with the data. Then, the same annotator annotates each of the tweets with an initial code that reflects the political message that the religious text is used to deliver, if such message clearly exists. Next, initial codes are grouped into initial themes that reflect the main tactics involved in delivering the political messages. The review process of the themes, which was conducted by the annotators in cooperation with other researchers, revealed that some of the themes are overlapping and that some samples involve multiple tactics that should be grouped into one theme. The annotator finally groups all the labels under three main themes, namely \texttt{Supporting one's own country or its allies}, \texttt{Criticism of opponents}, and \texttt{Non-political}. Table \ref{tab:codebook} shows the detailed description and categories of each theme. Finally, the second annotator annotates a random 5\% sample of the data. The inter-rater agreement achieved a Cohen's Kappa of 0.91 showing a significantly high alignment between annotators.}{change-R1C2}

\change{\subsection{Ethical Considerations}}{change-R2C5}

\change{Religion remains a sensitive discussion topic even in western communities \cite{schafer2018talks}. The authors of this work declare that discussing the negative impact of religion's political deployment is neither a denunciation of religion nor of religious groups. We highlighted the justification behind our choices of the datasets under study. Nonetheless, we believe that the practices reported in our study might be very likely applicable to other religions, communities, and religious groups which can be studied in future research work.}{change-R2C5}

\change{The Twitter Transparency Report \cite{twitter_transparency} is released in accordance with multiple regulations including the EU's Digital Services Act (DSA). The accounts inside the datasets released in this context are suspended before the release of the data; hence, the owners of those accounts cannot be reached or be subject to abuse as a result of the content they shared. We take additional data handling measures by ensuring the safe storage of processed and filtered data, hashing users IDs, and discarding sensitive attributes like screen names and profile descriptions. We also identified many tweets that mentioned personal accounts that do not belong to the dataset either to retweet their content or to reply to them so we removed those mentions and did not use these examples for demonstration. \\}{change-R2C5}

\section{Results}

\begin{table}[]
\begin{center}
\resizebox{0.48\textwidth}{!}{
\begin{tabular}{l c c c c c c} 
 \hline
  & KSA & UAE & Egypt & Turkey & Iran & Mixed \\
 \hline
 Quran Tweets  & 24,643 & 549  & 2,598 & 93 & 1,372 & 29,840 \\ 
 Verse Matches & 64,359 & 789  & 4,420 & 93 & 5,005 & 74,718 \\
 Hadith Tweets & 132,350 & 3,386 & 60,607 & 31 & 4,049 & 59,779  \\
 Hadith Matches & 113,775 & 2,646 & 55,049 & 31 & 949 & 45,968 \\
 \hline
\end{tabular}
}
\end{center}
    \caption{The number of tweets including Quran and Hadith in addition to the number of matches with Quran and Hadith instances.}
    \label{tab:religious-texts-counts}
\end{table}

\begin{figure}
\centering
\includegraphics[width=0.99\linewidth]{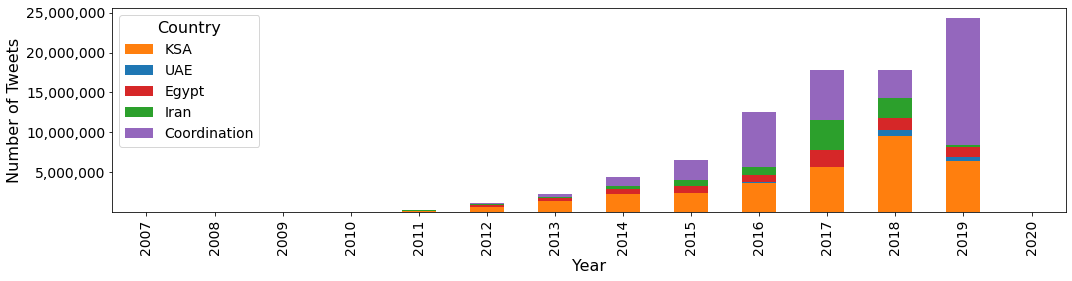}
\caption{\change{The distribution of all the tweets in the datasets per year.}{}}
\Description{The distribution of all the tweets in the datasets per year.}
\label{fig:time-all}
\end{figure}

\begin{figure}
\centering
\includegraphics[width=0.99\linewidth]{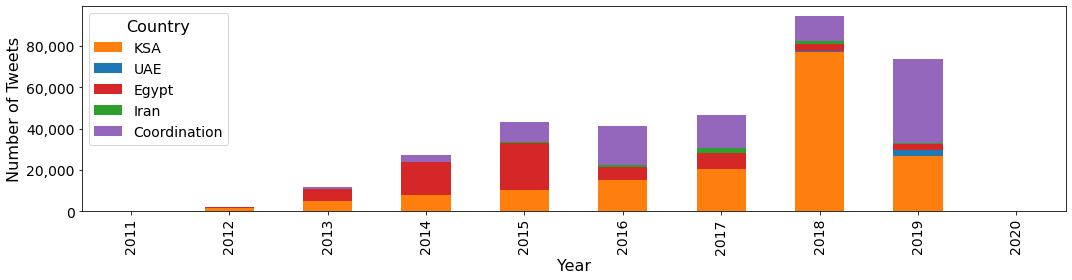}
\caption{\change{The distribution of the religious tweets in the datasets per year.}{}}
\Description{The distribution of the religious tweets in the datasets per year.}
\label{fig:time-religious}
\end{figure}

Table \ref{tab:religious-texts-counts} shows the presence of hundreds of thousands of tweets including Quran and Hadith, which confirms that state-sponsored propaganda accounts share this type of content. We notice that the number of tweets found for Turkey is very small so we discard it from our further analysis. This small number supports Twitter's comment about the Turkish dataset that it is mainly targeting a local audience. On the other hand, there are thousands of tweets including Arabic religious texts shared by the Iranian accounts which indicates  that it is partially targeting Arab audience. It is also worth noting that the number of verse matches is usually higher than the number of Quran tweets since a tweet part can match multiple verses as we mentioned earlier. In contrast, the number of hadith matches is less than the number of Hadith tweets since a tweet might include a hadith from the standard datasets or might include a text (possibly fabricated) that is not covered by them. \change{The distribution of all the propaganda tweets for each country per year shown in Figure \ref{fig:time-all} demonstrates that some of those accounts have been tweeting since 2007. The same distribution provided in Figure \ref{fig:time-religious} for religious tweets only shows that such texts could be detected starting from 2011 and that their volume per year does not always correlate with the overall volume of tweets in the same year suggesting that their presence is affected by events.}{} We analyze further the religious texts shared to examine their sharing patterns and check if they use religious texts to amplify any political messages.

\subsection{RQ1: Sharing Patterns}

Table \ref{tab:hadith-authenticity} shows the distribution of shared hadiths with respect to authenticity levels. A hadith can have one of four authenticity levels: \texttt{authentic}, \texttt{good}, \texttt{weak}, or \texttt{fabricated}. \citet{fawzi2025prophet} deal with \texttt{good} and \texttt{weak} as one category as they are both levels in the middle between \texttt{authentic} and \texttt{fabricated}. The \texttt{unmatched} row indicates the portion of tweets that had a Hadith keyword but was not matched against any of the hadiths in the datasets. Generally, all the distributions tend to be close to what is reported in the literature except for Iran's distribution where the percentage of fabricated hadiths and unmatched tweets are significantly higher. This is expected due to the disagreements between the two main branches of Islam namely, Sunni and Shia. The datasets used for evaluation are labeled based on the Sunni branch of Islam while Iran follows the Shia branch of Islam. This \change{bias}{change-R2C11} has also been reported by \citet{fawzi2026holiness} \change{and was mitigated in their study by involving a Shia scholar in the process. Nevertheless, the focus of our study is not Hadith fabrication but rather how Islamic texts in general are used for the purpose of propaganda and how different are the distributions in the propaganda datasets compared to the literature. Hence, we do not claim that these portions of fabricated or unmatched Hadith necessarily indicate planned disinformation}{change-R2C11}.

\begin{table}
\begin{center}
\resizebox{0.48\textwidth}{!}{
\begin{tabular}{l c | c c c c c} 
 \hline
  & \textbf{Literature} & KSA & UAE & Egypt  & Iran & Mixed \\
 \hline
 Authentic & \textbf{53.1}\% & 76.8\% & 66.5\%  & 83.3\%  & 10.3\% & 65.4\%\\ 
 Good/Weak & \textbf{13.9\%} & 8.6\% & 10.4\%  & 7.2\%  & 4.7\% & 10.1\%\\ 
 Fabricated & \textbf{1\%} & 0.6\% & 1.2\%  & 0.3\%  & 8.3\% & 1.4\%\\ 
 Unmatched & \textbf{32\%} & 14\% & 21.9\%  & 9.2\%  & 76.7\% & 23.1\%\\ 
 \hline
\end{tabular}
}
\end{center}
    \caption{The authenticity distribution of Hadith per data collection.}
    \label{tab:hadith-authenticity}
\end{table}

\begin{figure*}
\centering
\includegraphics[width=0.99\linewidth]{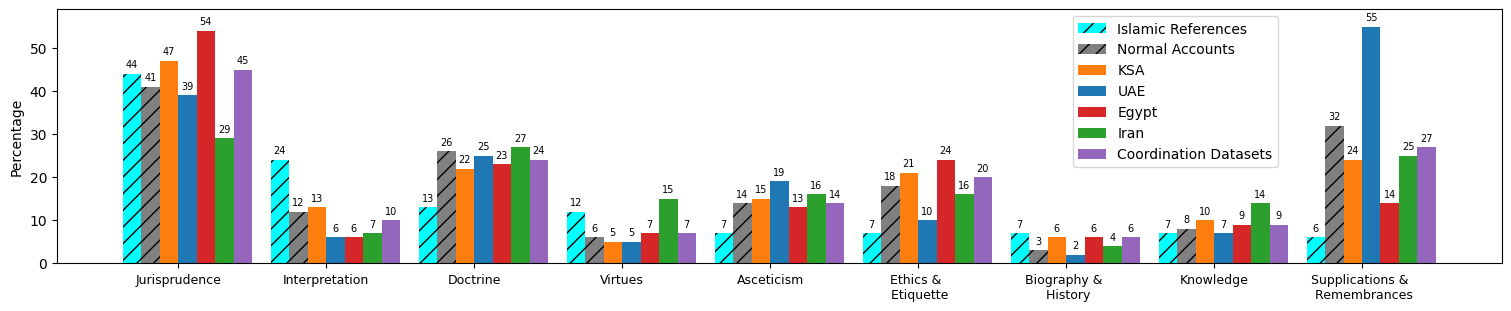}
\caption{The topical distribution of shared Hadith by different networks compared to \change{the distribution in Islamic references and}{} the distribution of normal accounts reported by \citet{fawzi2025prophet}.}
\Description{The topical distribution of shared Hadith by different networks compared to the distribution in Islamic references and the distribution of normal accounts reported by \citet{fawzi2025prophet}.}
\label{fig:topical-hadith}
\end{figure*}

\begin{figure*}
\centering
\includegraphics[width=0.99\linewidth]{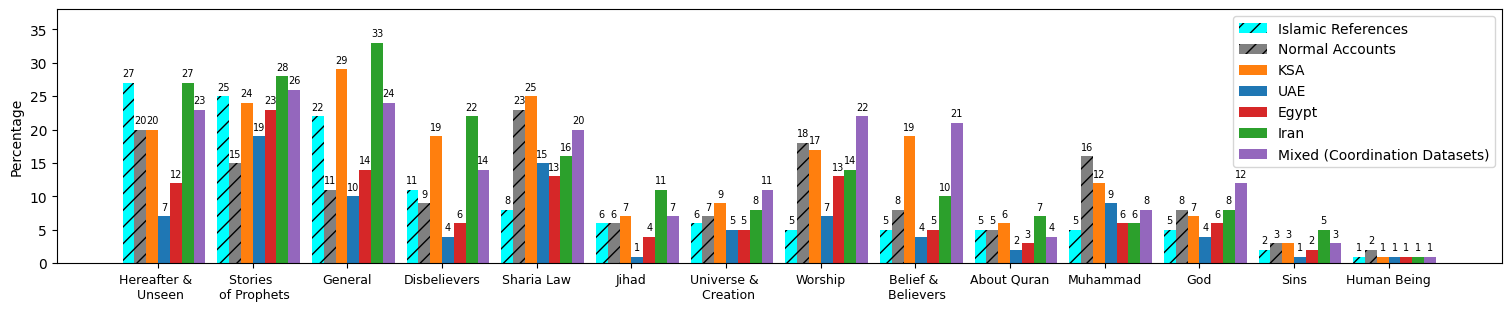}
\caption{The topical distribution of shared Quran by different networks compared to \change{the distribution in Islamic references and}{} the distribution of normal accounts reported by \citet{abokhodair2020holy}.}
\Description{The topical distribution of shared Quran by different networks compared to \change{the distribution in Islamic references and} the distribution of normal accounts reported by \citet{abokhodair2020holy}.}
\label{fig:topical-quran}
\end{figure*}

We investigated what topics  the hadiths and Quran quotes shared by the accounts were about. To categorize them into topics, we follow the existing classifications by the two websites \texttt{sonnaonline}\footnote{Reference for Hadith topics: \url{sonnaonline.com}} and \texttt{quranbysubject}\footnote{Reference for Quran topics: \url{quranbysubject.com}}. It is important to note that a hadith or a Quran verse can belong to multiple categories; hence, the sum of the percentages of categories can exceed 100\%. The topical distributions of hadiths in Figure \ref{fig:topical-hadith} show that the category of \textit{Jurisprudence} is overrepresented in the Saudi, Egyptian, and mixed data collections. This category covers how Islam should be applied within religious practice, private life, and public life. On the other hand, the Emirati data feature an over-representation for the topics \textit{Asceticism} and \textit{Supplications \& Remembrances} which call for paying less attention to materialistic life and more attention to spirituality. Finally for Iran, there is an increase in the portion of the \textit{Virtues} category. This category has a special importance in the Shia branch of Islam since it involves praising multiple characters in the Islamic history like the cousin of the prophet, Ali, and Ali's two sons, Alhassan and Alhussein. The worthiness of these characters of political leadership is one of the main disagreement points between Sunnis and Shias that keep getting discussed until today. Figure \ref{fig:topical-quran} presents the topical distributions for Quran. The distributions of KSA, Iran, and the coordinated datasets (which include KSA) show a higher presence of multiclass verses compared to Egypt and the UAE. This is evident from their over-representation of most categories, whereas Egypt and the UAE tend to under-represent most categories. Multiclass verses are generally more complex, which allows them to fit into multiple categories and deliver multiple meanings simultaneously. All collections have higher portions than the literature for the category of \textit{Stories of Prophets} which shows general reliance on it. KSA and Iran considerably emphasize the topic of \textit{Disbelievers}, which is a negative description in the Quran that can potentially be used to criticize others as we show later. While Iran also emphasizes \textit{Jihad} and \textit{Hereafter \& Unseen}, KSA 
emphasizes \textit{Belief \& Believers} (that can be used for praise as we show later), which is also emphasized by the coordinated datasets along with \textit{Worship} and \textit{God}.

Next, we explore the common themes among the top shared Islamic texts per country. The full listing of the top shared texts can be found in appendix \ref{appendixQuran} and appendix \ref{appendixHadith}. To facilitate referring to those texts we use the following notation $Tn_{CA}$ where \texttt{T} is the type of the text (Quran "Q"/Hadith "H"), \texttt{n} is its ranking index, \texttt{C} is the first letter of the country name, and \texttt{A} is its authenticity (only applies for Hadith).

\begin{itemize}
    \item \textbf{KSA:} The top religious texts shared in the Saudi collection resemble the top ones typically shared with few exceptions. The top shared Quran verse $Q1_{K}$ reads \textit{We certainly know that your heart is truly distressed by what they say.} with some instances accompanying it with the hashtag \<درع سلمان للدعم>\# (\#Salman's\_support\_shield), showing support to King Salman against critics. $Q3_{K}$ reads \textit{Any slave whom Allah makes him in charge of subjects and he dies while he is not sincere to them, Allah will make Jannah unlawful for him}, which tells how big the responsibility is to be in charge of people as a ruler.
    \item \textbf{UAE:} The top Quran verses $Q1_{U}$ and $Q2_{U}$ show spiritual optimism far from action as they read \textit{So, surely with hardship comes ease.} and \textit{And Allah is fully capable of doing all things}. This confirms the same impression observed within UAE's sharing patterns. Most of the top hadiths shared in the Emirati dataset belong to the \textit{Supplications \& Remembrances} category, confirming the same impression. We find two instances of sharing the following hadith: \textit{The Persian language is most hated talk to God.} which demonstrates a hate sentiment towards Iran.
    \item \textbf{Egypt:} The top Quran verse shared reads \textit{And remember when Job cried out to his Lord, “I have been touched with adversity, and You are the Most Merciful of the merciful.”} which calls for patience. Egypt has been undergoing tough economic conditions in recent years \cite{sayigh}. $Q3_{U}$, which reads \textit{There is security of life for you in the law of retaliation, O
    people of reason, so that you may become mindful of Allah.}, has been used to call for revenge and imprisonment of all Islamist leaders so that they become an example to others. The verse has been linked with the hashtag \<مصر مش هتضيع يا إخوان> \# (O Muslim Brotherhood, Egypt is not going to fall).  For Hadith, $H1_{Ea}$ is the most shared and it reads \textit{Among the people who will be most severely punished on the Day of Resurrection will be the image-makers.} While Muslim scholars interpret image-makers in this hadith as people creating sculptures, it can also be misused to mean photographers. Multiple other hadiths urge people not to mention negative sides of others and not dig deep into issues and keep analyzing them. For example, $H1_{Egw}$ reads \textit{Allah dislikes an eloquent person who rolls his tongue as a cow rolls its tongue} and $H2_{Egw}$ reads \textit{None of my companions should tell me anything about anyone, for I would like to come out to you with a sound heart}. These texts align with the reported restriction of freedom of speech and protest mentioned by \citet{springborg2025egypt}.
    \item \textbf{Iran:} Two main themes arise in Iran's top texts: asserting the strong connection between Muslims beyond nations (e.g. $H1_{Ia}$ and $H5_{Ia}$) in addition to promoting the concepts, central characters, and worthiness of leadership of Shia political Islam (e.g. $Q2_{I}$, $Q3_{I}$, $H1_{Igw}$ and $H3_{Igw}$). Some of these hadiths are among the top fabricated hadiths found to be shared in earlier works (e.g. $H2_{If}$ and $H3_{If}$) \cite{fawzi2026holiness}. $H1_{Ia}$ which promotes the first theme reads \textit{A faithful believer to a faithful believer is like the bricks of a wall, reinforcing each other.} This aligns with Iran's offshore support to multiple military and political groups like Houthi in Yemen and Hezbollah in Lebanon. It justifies the intervention in the politics of other countries through this support.
    \item \textbf{Mixed (Coordination Datasets):} The top two Quran verses urge people to donate money for the poor and have been shared along with donation campaigns. The first one reads \textit{and give food despite their desire for it to the poor, the orphan, and the captive.} and the second reads \textit{Whatever donations you give are for parents, relatives, orphans, the poor, and needy travelers}. For Hadith, the only interesting hadith we found is fabricated and reads \textit{Eggplant is a cure for every disease.}, which is medical misinformation that has circulated during the time of COVID-19.
\end{itemize}

\subsection{RQ2: Political Messages}

\begin{table}[]
\begin{center}
\begin{tabular}{l c c  c c c} 
 \hline
  & KSA & UAE & Egypt  & Iran & Mixed \\
 \hline
 Quran Tweets & 417 & 36 & 358 & 141 & 608 \\ 
 Hadith Tweets & 878 & 23 & 763 & 120 & 629 \\
 \hline
\end{tabular}
\end{center}
    \caption{The number of tweets including Quran and Hadith after filtering the collections using the keywords in table \ref{tab:political-keywords}}
    \label{tab:political-texts}
\end{table}

\begin{figure}
\centering
\includegraphics[width=\linewidth, height=6cm]{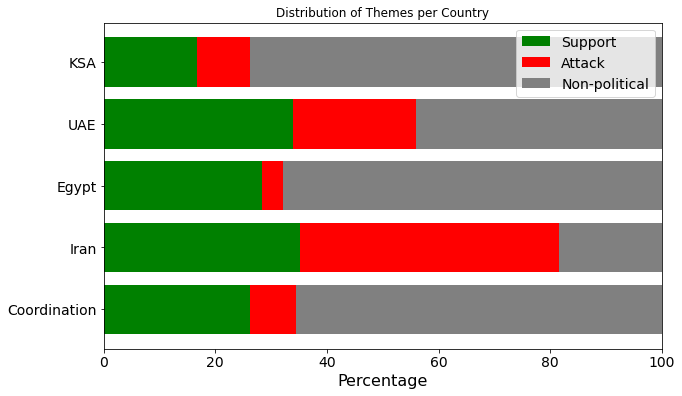}
\caption{\change{The distribution of the tweets with political keywords per theme.}{}}
\Description{The distribution of the tweets with political keywords per theme.}
\label{fig:rq2-statistics}
\end{figure}

Twitter reports that high volumes of the tweets in some datasets are shared to mask manipulation. By filtering the tweets using keywords representing political entities, we take a closer look at the general political messages of different networks. Table \ref{tab:political-texts} shows the number of tweets extracted per data collection; however, those numbers are not representative of the portion of political tweets as some of the keywords already exist in religious texts like \textit{Egypt}, \textit{Qatar} (with another meaning), \textit{Yemen}, \change{and \textit{Salman} (referring to \textit{Salman the Persian}, who is a religious Islamic scholar and a companion of prophet Muhammad, not the current king of Saudi Arabia). Figure \ref{fig:rq2-statistics} shows the thematic distribution per country that we found after going through \textbf{all} the 3,973 tweets having the keywords in table \ref{tab:political-keywords}. While a big portion of those tweets is \texttt{non-political} as per the definition in table \ref{tab:codebook}, there are big portions of \texttt{support} tweets across all countries. It is also worth noting that UAE and Iran have a much bigger portion of \texttt{Attack}. This is expected for both countries given the reports about their offshore intervention especially Iran whose local language is Farsi; hence its Arabic content is expected to be directed to an offshore audience}{}. We demonstrate some examples for each country in table \ref{tab:political-tweets} and we discuss them in the rest of this section. \change{The original Arabic tweets for these examples can be found in Appendix \ref{sec:political-arabic}.}{}
\begin{table*}[]
\begin{center}
\small \begin{tabular}{c p{13.5cm} c c c} 
 \hline
 \# & Tweet (translated) & Country & R. & Th. \\
 \hline
  1 & Dr. Abdullah Al-Shathri: \#King\_Salman directed the launch of \#Operation\_Decisive\_Storm in fulfillment of the Almighty’s saying: \textbf{“if they seek your help in faith against persecution, it is your obligation to help them”} & KSA & Q & S \\
  2 & God Almighty said: \textbf{"Whenever We intend to destroy a society, We command its elite to obey Allah but they act rebelliously in it."} \#Hamad\_Organization \#Down\_with\_the\_Qatari\_Terrorist\_Regime \#Qatar & KSA & Q & A \\
 3 & The Messenger of Allah (peace and blessings be upon him) said: \textbf{"Two eyes will never be touched by the fire of Hell; an eye which weeps out of Fear of Allah and an eye which spends the night in guarding in the Cause of Allah"} \#OurSoldiersAreBombingIranianMilitias & KSA & H & S \\
 4 & O Iraq, you will remain proud despite the machinations of the spiteful. The Messenger of God, peace and blessings be upon him, said: \textbf{When Khosrow is ruined, there will be no Khosrow after him.} Die in your rage, no matter what you do, for victory is ours. This is a promise from God. With my regards, a lover of Iraq and all Islamic countries. \#Iraq\_is\_being\_subjected\_to\_Iranian\_genocide & KSA & H & A \\
 5 & Unfortunately, there isn't a single man in \#Qatar who can stop Bint Awais after she declared God's limits to be terrorism. God Almighty said: \textbf{"Indeed, the penalty for those who wage war against Allah and His Messenger and spread mischief in the land is death, crucifixion, cutting off their hands and feet on opposite sides, or exile from the land. This penalty is a disgrace for them in this world, and they will suffer a tremendous punishment in the Hereafter."} & UAE & Q & A \\
 6 & God Almighty said: \textbf{(And do not be inclined to the wrongdoers or you will be touched by the Fire....)} Do not trust or rely on the tyrants and oppressors, those who wield power on earth and subjugate people with their might. Do not rely on them. \#SudanCitiesRevolt \#DownWithTheRegime & UAE & H &  A  \\
 7 & \textbf{Never think of those martyred in the cause of Allah as dead. In fact, they are alive with their Lord, well provided for.} God Almighty has spoken the truth. May God have mercy on them, forgive them, grant them spacious gardens in Paradise, and inspire their families with patience and solace. Indeed, we belong to God, and to Him we shall return. We are with you, Mr. President El-Sisi, and with our great and valiant army and our great police force. Intellectual terrorism must be confronted. @AlsisiOfficial & Egypt & Q & S \\
 8 & Summary of what happened in 
 \#Turkey..God Almighty said: \textbf{“And Allah drove back the disbelievers in their rage, totally empty-handed.} And Allah spared the
believers from fighting. For Allah is All-Powerful, Almighty.” & Egypt & Q & A  \\
9 & The Messenger of God said, \textbf{"If God opens Egypt for you, then establish a large army there, for that army is the best army on earth."} \#OurEgyptianIdentityIsStrongerThanYou & Egypt & H & S \\
10 & \#WeAreStrongerThanYourExplosions \textbf{When you conquer Egypt, treat its inhabitants well. For there lies upon you the responsibility because of blood ties with them.} \#UnitedOnPalmsunday  & Egypt & H & S \\
11 & \#TheNewSaudiArabiaILikeIt because it will be the shortest country in history, and the proof is the Almighty's saying: In the name of God, the Most Gracious, the Most Merciful: \textbf{Whenever We intend to destroy a society, We command its elite to obey Allah but they act rebelliously in it.} & Iran & Q & A \\
12 & \#BinSalmanIsStarvingYemen God Almighty said: \textbf{(Verily, when kings enter a town, they destroy it and make the most honorable amongst its people the lowest.)} This is the state of the Al Saud family in Yemen. & Iran & Q & A \\
13 & 
The Messenger of God (peace and blessings be upon him and his family) said, \textbf{“Whoever pleases an unjust ruler by angering God has left the religion of God.”} This applies to all those who S the aggression against Yemen & Iran & H & A \\
14 & The Messenger of God (peace be upon him) said: \textbf{“O people, spread peace, feed the hungry, maintain ties of kinship, and pray at night while people are asleep, and you will enter Paradise in peace.”} Bin Salman has cut off food and severing ties of kinship, and has spread killing and destruction! \#BinSalmanIsStarvingYemen & Iran & H & A \\
15 & God willing, you will return to \#Aleppo to liberate it from the gangs of \#Assad and Iran. Just act according to the words of God Almighty: \textbf{[And hold firmly together to the rope of Allah and do not be divided.]}. & Coordination & Q & S \\
16 & \#QatarConsidersIranAnHonorableState  Yes, it is honorable! But only in the eyes of traitors. The Messenger of God, peace and blessings be upon him, said: \textbf{((By Allah, he is not a believer. One whose neighbor does not feel safe from his evil))}'
 & Coordination & H & A \\
 17 & God Almighty said: \textbf{"Did We not soothe your heart for you?"} O God, by the depth of this verse, expand our hearts, comfort our souls, and remove our worries. Indeed, You are capable of all things." \#Qatar
 & Coordination & Q & NP \\
 18 & The Messenger of Allah (peace and blessings be upon him) said, \textbf{"Patience is half of faith, and fasting is half of patience."} \#BlessedRamadan \#UAE
 & Coordination & H & NP \\

\hline
\end{tabular}
\end{center}
    \caption{\change{Examples of tweets using religious texts to deliver political messages (original Arabic version in Appendix \ref{sec:political-arabic}). R.: Religious text in the tweet, Q: Quran, H: Hadith, Th.: Theme, S: Support, A: Attack, NP: Non-political.}{}}
    \label{tab:political-tweets}
\end{table*}

\subsubsection{KSA} 

\change{The majority of the Saudi tweets are non-political, but the political ones focus on supporting the regime and its allies more than attacking opponents.}{} Saudi propaganda shares a lot of Quran verses and Hadith praising Egypt, which shows the alliance between the two states during this period. Another theme is justifying \textit{Operation Decisive Storm} which is the Saudi-led intervention in the Yemeni civil war since March 2015. \change{Example \#1}{}  shows a Quran verse reading \textit{if they seek your help in faith against persecution, it is your obligation to help them} along with support for the King's decision to start the operation. \change{Example \#3}{}  shows a hadith saying \textit{Two eyes will never be touched by the fire of Hell; an eye which weeps out of Fear of Allah and an eye which spends the night in guarding in the Cause of Allah}, followed by reporting that Saudi soldiers are bombing Iran's militia. The eye which spends the night in guarding in the Cause of Allah refers to soldiers fighting in the righteous Islamic army, in this case, the Saudi army. Finally, the third prevalent theme is attacking enemies like Iran and Qatar. For example, the hadith in \change{example \#4}{}  reads \textit{When Khosrow is ruined, there will be no Khosrow after him}. Khosrow is the last Sasanian king who ruled Iran and the hadith symbolized that Iran will not be a great empire again. The author follows the hadith by showing support for Iraq against the Iranian violence there. The same strategy was used against Qatar as in \change{example \#2}{}  which includes the Quran verse \textit{Whenever We intend to destroy a society, We command its elite to obey Allah but they act rebelliously in it. So the decree of punishment is justified, and We destroy it utterly}. Many tweets include this verse then accuse the Qatari regime of corruption and criticize the amount of money they spent for organizing the 2022 FIFA World Cup.

\subsubsection{UAE}

\change{Besides praising own country and leaders,}{} the general direction of the Emirati tweets is criticizing offshore targets. \change{Example \#6}{} includes a verse that reads \textit{And do not be inclined to the wrongdoers or you will be touched by the Fire.} This is followed by a call for the Sudanese people not to be inclined to their government as they oppress them. The Emirati intervention in Sudan has continued even after the fall of Omar al-Bashir. In March 2025, Sudan has filed a case against UAE in the International Court of Justice (ICJ) accusing it of being complicit in genocide inside Sudan\footnote{Sudan's case against UAE: \url{icj-cij.org/case/197}}. The other example in \change{example \#5}{} criticizes Qatar for not punishing Aljazeera's Christian journalist, Ghada Owais, after she -- as they claim -- attacked Islam. The attached Quran verse reads \textit{Indeed, the penalty for those who wage war against Allah and His Messenger and spread mischief in the land is death, crucifixion, cutting off their hands and feet on opposite sides, or exile from the land. This penalty is a disgrace for them in this world, and they will suffer a tremendous punishment in the Hereafter}.

\subsubsection{Egypt} 

Egyptian \change{support}{} propaganda focuses on praising the efforts exerted by the police and the army to counter terrorism and showing support to them. \change{The attack tweets are relatively much less.}{} The author in \change{example \#7}{} adds the Quran verse \textit{Never think of those martyred in the cause of Allah as dead. In fact, they are alive with their Lord, well provided for} and follows it with expressing support for the Egyptian president, the army, and the police against terrorism. The hadith in \change{example \#9}{} reads \textit{If God opens Egypt for you, then establish a large army there, for that army is the best army on earth.} and is followed by a hashtag whose translation is \#OurEgyptianIdentityIsStrongerThanYou. In addition, they show solidarity with Egyptian Christians against incidents of terrorism performed against them. \change{Example \#10}{} starts with a hashtag whose translation is \#WeAreStrongerThanYourExplosions and then includes the hadith \textit{when you conquer Egypt, treat its inhabitants well. For there lies upon you the responsibility because of blood ties with them}. The inhabitants of Egypt before the Islamic rule were Christians. Finally, they also dedicate a lot of tweets for criticizing entities linked with political Islam like Turkey, Qatar, and the government of western Libya as well as praising allies like the Emirati armed forces. \change{Example \#8}{} says that the summary of what happened in Turkey (during the 2016 coup attempt) is the following verse \textit{And Allah drove back the disbelievers in their rage, totally empty-handed. And Allah spared the believers from fighting. For Allah is All-Powerful, Almighty.} referring to the Turkish regime as the disbelievers.

\subsubsection{Iran} The main direction of Iran's propaganda is supporting its Shia political allies across the Middle East like Hezbollah in Lebanon and Houthis in Yemen in addition to criticizing their enemies especially Saudi Arabia. \change{Examples \#11}{} and \change{\#14}{} criticize the destruction caused by the intervention of Saudi Arabia in Yemen. The Quran verse reads \textit{Verily, when kings enter a town, they destroy it and make the most honorable amongst its people the lowest.}. The author of the tweet says then that this applies to Saudis in Yemen and uses the hashtag \<بن سلمان يجوع اليمن>\# (\#BinSalmanIsStarvingYemen). The same hashtag is used with the hadith that reads \textit{O people, promote the greetings, feed (the poor and needy) and perform prayers when others are asleep so that you will enter heaven safely} and is followed by claiming that Muhammad Bin Salman, KSA's Crown Prince, has cut off food supplies to Yemen and spread murder and destruction. The tweet in \change{example \#13}{} criticizes supporters of attacking Yemen and attaches the hadith \textit{Whoever pleases an unjust ruler by angering God has left the religion of God.} Interestingly, \change{example \#12}{} uses the same Quran verse that the Saudis used to criticize Qatar in \change{example \#2}{} to criticize Saudi Arabia. It precedes the verse by the claim that the new Saudi Arabian state will be the shortest living state in history due to the corruption of its government.

\subsubsection{Coordination Datasets} The themes in this data collection are similar to the ones found in the individual propaganda of the coordinating countries. Essentially, they include promoting their narrative in countries like Libya, Sudan, and Syria in addition to praising the regimes in KSA, UAE, and Egypt. The author of \change{example \#15}{} tells Syrians that they shall return to Aleppo (a Syrian city) and free it from Al-Assad's gangs (the former Syrian president who was an ally to Iran) and they just have to stick to the Quran verse \textit{And hold firmly together to the rope of Allah and do not be divided}. The author in Figure \change{example \#16}{} starts the tweet with the hashtag \#QatarConsidersIranAnHonorableState and then continues to say that indeed Iran is honorable but only in the eyes of traitors. Finally, the tweet ends with the hadith \textit{By Allah, he is not a believer. One whose neighbor does not feel safe from his evil} referring to Iran.

\section{Discussion}
The objective of our study was to answer two main research questions: (RQ1) How do propagandists’ patterns of sharing religious texts differ from standard sharing patterns, if they share such texts at all? and (RQ2) Do propagandists use religious texts to promote their agendas, and how? After processing around 124 million tweets \change{from 32.5K accounts}{}, our results confirm that propagandists in the Middle East share high volumes of Islamic texts. Their sharing patterns in terms of topical and authenticity distributions align with their state's nature. By manually investigating around 4,000 tweets, we observe that propagandists clearly use such texts to serve their agendas. The tactics involve praising their own government and leadership, criticizing countries, celebrities, and parties who are considered opponents, interpreting past events and predicting future events based on interpretations of religious texts, and justifying political decisions by relating them to religious texts. \change{These results have important implications for the HCI community that we discuss below.}{}

\subsection{\coloring{Distortion of Online Islamic Content by State-sponsored Propaganda}}
\change{We aim through our findings to add to the existing research directed towards designing better social media platforms \cite{hartmann2025lost, meinhardt2025scrolling}. Recent HCI works have covered considerations related to a diverse set of groups like adolescents \cite{fischer2025building}, activists \cite{luther2025social}, and users with disabilities \cite{heung2025ignorance, sharevski2025have}. We extend this by covering an aspect related to another massive yet understudied group, namely Muslims. Social studies discuss how the desire of influencers to maximize their reach by sharing Islamic texts impacts how they select and frame them in ways that do not maintain their theological integrity \cite{simamora2024rethinking, kaplan2023religious}. Our study shows that this threat is also present on a larger scale when religion is deployed by states for political manipulation. In this context, state-sponsored propaganda is an additional challenge for online Islamic content causing its distortion. This effect is a manifestation of how online religious practice is influenced by the inherent side effects of social media like hate speech \cite{castano, albadi2019hateful}, misinformation \cite{alimardani2020covid19}, and polarization \cite{al2021atheists}. This is indeed an effect that can deteriorate the experience of religious groups on such platforms.}{change-R1C1}

\subsection{\coloring{Digital Propaganda Targeting Muslim Users}}

\change{HCI research still lacks works that attempt to deeply understand religious communities \cite{wolf2024still, buie2013spirituality}. In fact,}{}  another contribution of this study is shedding light on a rich collection of understudied datasets that is almost never considered by HCI researchers. Generally, these datasets reveal a lot about the political and the cultural dynamics in the countries involved in them as well as the countries targeted or supported by the involved ones. A recent study reports that \change{there appear to be similarities between what is considered hateful on Persian and on Arabic social media}{}, particularly in targets related to religion, politics, and common controversial events that provoke hate \cite{bokaei-etal-2025-culture}. While the claim is supported by linguistic experiments, our study provides social explanation for it through showing the Iranian involvement in Arab politics and the other way around. \change{It also highlights the Middle East as a theater where religion as a form of ideology serves as a weapon used by the elite power to achieve its goals and interests \cite{hyzen2021revisiting}. The variation in the religious texts used per country and per situation is also a practical supporting example to the digital propaganda theories suggesting that propaganda is changing substantively in reaction to major developments \cite{boussalis2023matters}.}{} Finally, our study questions the plausibility of the attempts for secularizing Arab societies in their current state. While some regional powers adopt this agenda \cite{davidson2019uae}, there are structural issues in their policies like sharp power interventions that cause inconsistencies with this theoretical framework \cite{kourgiotis2020moderate}. Our findings show that suppressing political Islamist opposition groups merely shifts religious political deployment into the hands of autocratic states, rather than diminishing it.

\subsection{\coloring{Religion as a Tool for Online Manipulation}}
Our study is, to the best of our knowledge, the first work that quantitatively examines the phenomenon of the deployment of Quran and Hadith for propaganda in the Middle East, a highly volatile and religious yet understudied region. The examination of the understudied theme of religious propaganda on social media platforms contributes to the research direction that is concerned with specialized forms of manipulation on such platforms. While many works within moderation research still focus on classical forms of manipulation like political misinformation \cite{fawzi2024pinocchio}, in response to global events, some works stepped in to investigate unusual forms like medical misinformation \cite{juneja2021auditing}, corporate misinformation \cite{zhou2024characterizing}, and religious misinformation \cite{fawzi2026holiness}. By that, the HCI community gains new insights that help replace outdated moderation strategies like censorship that started to be ineffective \cite{ccetinkaya2025state, hakami2025cooperative}. Supporting users \change{and raising their awareness}{} need a deeper understanding of aspects like how conflict detection and selective avoidance work \cite{tanaka2023does}. Our work \change{suggests that}{} religion intervenes in this process as people naturally tend to avoid information that contradicts their preexisting beliefs \cite{fischer2005selective}. It also \change{suggests that}{} religion is among the aspects that sometimes prevent people from investigating misinformation. \citet{geeng2020fake} observed personal multiple reasons why people skip fact-checking misinformation. Although their dataset of fake threads included a religious sample with the title "Church", they did not discuss the effect of religion. Some qualitative studies suggest the presence of this effect in the west \cite{pennycook2016atheists, bronstein2019belief}; \change{hence,}{} targeted moderation could be effective in the same way targeted moderation towards electoral misinformation was \cite{juneja2023assessing}. 

\subsection{\coloring{Recommendations for Designing More Religion-Friendly Social Media Platforms}}

\change{We address social media platform designers as the audience of our paper. To achieve targeted moderation towards religious texts, social media platforms are encouraged to take the following measures:}{}

\change{\begin{enumerate}
    \item Cooperate with religious scholars to set standard measures for these texts \cite{simamora2024rethinking}. Islamic scholars have recently participated in multiple initiatives that aim to make online religious content more accurate \cite{bouchekif2025qias, mubarak2025islamiceval} driven by the rapid development in LLMs.
    \item Use analytics to monitor how religious texts interact with ongoing trends. For this, more advanced algorithms for religious texts extraction need to be developed and used \cite{kamran2024semantic}. In particular, these algorithms need to be context-aware \cite{sahebi2025context}.
    \item  Promote trusted religious online resources as well as communities \cite{simamora2024rethinking}. This approach has also been suggested for rumors and fake news where relevant authorities are invited for validation \cite{haouari2023can}, so the same approach could be extended to religious content.
\end{enumerate}}{}

\change{This adaption will not only improve the experience of the users in the Middle East but rather many users who have been negatively impacted by its absence in other parts of the world \cite{Biswas2025SocialMediaReels, rifat2024politics, bowman2025sharing}.}{}

\subsection{Limitations \& Future Work}
We aim to study how state-sponsored social media accounts share and use religious texts to further their goals, but due to the data available in the datasets shared by Twitter’s transparency reports, our efforts are limited to tweets about Islamic texts in Arabic.  This is a limitation in several ways: one is that only a minority of Muslims converse daily in Arabic. Quran and Hadith are exchanged by hundreds of millions of Muslims in other languages \cite{fawzi2026holiness}, and our study does not look at Muslim populations in other Muslim-majority countries such as Indonesia or Pakistan \change{which do not have similar datasets}{}. The second is that we do not look at how religions other than Islam are harnessed to further the political aims of governments, for example Christianity by Russia and the US or Hinduism by India. \change{Furthermore, since these datasets do not provide comprehensive details about the interaction networks of these accounts like their follow/retweet/reply networks, an inspection of the propagation or these religious texts and the reactions of users who are not propagandists towards them was not possible.}{change-R3C8, R3C5}

Our study is limited to the text of tweets. The Twitter data includes many photos and videos among which we found examples of religious propaganda \change{like Figure \ref{fig:examples}}{}. This would be worth studying in future research, and it would be interesting to explore how present religious propaganda is on other platforms beyond Twitter (X). 

Finally, we are limited to the state-sponsored accounts and coordinated information campaigns that Twitter chose to identify and publicly release in their Transparency Report. As a for-profit company based in the US, their choices will naturally have been influenced by a variety of considerations that we do not wish to speculate on here, but as a result the data may not include all accounts used for state-sponsored propaganda during the time frame under consideration here. Still, despite its limitations, this is tremendously useful data for academic research, and we encourage social media platforms to repeat this kind of project and grant access to data to researchers so they can perform this vital type of study.

\section{Conclusion}
This study provides an initial understanding for the interplay between religion and propaganda on social media platforms in Arabic. It shows that propagandists share religious texts with high volumes and odd patterns potentially affecting the overall religious presence on such platforms. It also shows that religion provides propagandists with a toolkit of tactics to serve their agendas either by criticizing their opponents and/or supporting their allies. This connection raises a lot of questions about social media users, content moderation, theories of propaganda, and the cultural as well as the political topology of the region of the Middle East. We hope that this work will serve as a decent building block for the HCI community and will contribute to new design considerations of social media platforms that were not present before.

\begin{acks}
We thank Dr. Savvas Zannettou and Dr. Tuğrulcan Elmas for their assistance with data access. We also thank Dr. Dilara Keküllüoğlu for her helpful discussions and feedback. We dedicate this work to the victims of propaganda and violence in the countries involved in our study, whose experiences and losses motivate our continued efforts to inspect new techniques that can reduce this harm.
\end{acks}

\bibliographystyle{ACM-Reference-Format}
\bibliography{sample-base}

\appendix

\section{Top shared Quran}
\label{appendixQuran}
\FloatBarrier

\begin{table}
\centering
\begin{tabular}{ p{1cm} R{7cm}}
\toprule
\multicolumn{1}{l}{\textbf{Count}} & \textbf{Top Quran Verses/Fragments (KSA)}\hspace{0cm}\mbox{ } \\ 
\midrule
1673 & \<وَلَقَدْ نَعْلَمُ أَنَّكَ يَضِيقُ صَدْرُكَ بِمَا يَقُولُونَ> \\
& We certainly know that your heart is truly distressed by what they say. \\
962 & \<يَا أَيُّهَا الَّذِينَ آمَنُوا اذْكُرُوا اللَّهَ ذِكْرًا كَثِيرًا> \\
& O believers! Always remember Allah often \\
\bottomrule
\end{tabular}
\caption{The top shared Quran verses in the data collection of KSA along with their translation from \url{quran.com}. For some long verses, only the fragments commonly shared in tweets are added.}
\label{tab:top_quran_ksa}
\end{table}

\begin{table}
\centering
\begin{tabular}{ p{1cm} R{7cm}}
\toprule
\multicolumn{1}{l}{\textbf{Count}} & \textbf{Top Quran Verses/Fragments (UAE)}\hspace{0cm}\mbox{ } \\ 
\midrule
47 & \<فَإِنَّ مَعَ الْعُسْرِ‌ يُسْرً‌ا> \\
& So, surely with hardship comes ease. \\
37 & \<وَكَانَ اللَّهُ عَلَىٰ كُلِّ شَيْءٍ مُّقْتَدِرًا> \\
& And Allah is fully capable of doing all things. \\
\bottomrule
\end{tabular}
\caption{The top shared Quran verses in the data collection of UAE along with their translation from \url{quran.com}. For some long verses, only the fragments commonly shared in tweets are added.}
\label{tab:top_quran_uae}
\end{table}

\begin{table}
\centering
\begin{tabular}{ p{1cm} R{7cm}}
\toprule
\multicolumn{1}{l}{\textbf{Count}} & \textbf{Top Quran Verses/Fragments (Egypt)}\hspace{0cm}\mbox{ } \\ 
\midrule
152 & \<وَأَيّوبَ إِذ نادى رَبَّهُ أَنّي مَسَّنِيَ الضُّرُّ وَأَنتَ أَرحَمُ الرّاحِمينَ> \\
& And remember when Job cried out to his Lord, “I have been touched with adversity, and You are the Most Merciful of the merciful.” \\
129 & \<وَهِيَ تَجْرِي بِهِمْ فِي مَوْجٍ كَالْجِبَالِ> \\
& And so the Ark sailed with them through waves like mountains. \\
87 & \<وَلَكُمْ فِي الْقِصَاصِ حَيَاةٌ يَا أُولِي الْأَلْبَابِ لَعَلَّكُمْ تَتَّقُونَ> \\
& There is security of life for you in the law of retaliation, O people of reason, so that you may become mindful of Allah.
 \\
\bottomrule
\end{tabular}
\caption{The top shared Quran verses in the data collection of Egypt along with their translation from \url{quran.com}. For some long verses, only the fragments commonly shared in tweets are added.}
\label{tab:top_quran_egypt}
\end{table}

\begin{table*}
\centering
\begin{tabular}{ p{1cm} R{16cm}}
\toprule
\multicolumn{1}{l}{\textbf{Count}} & \textbf{Top Quran Verses/Fragments (Iran)}\hspace{0cm}\mbox{ } \\ 
\midrule
36 & \<لَعَلَّ اللَّهَ يُحْدِثُ بَعْدَ ذَلِكَ أَمْرًا> \\
& perhaps Allah will bring about a change of heart later \\
26 & \<أَفَحُكْمَ الْجَاهِلِيَّةِ يَبْغُونَ ۚ وَمَنْ أَحْسَنُ مِنَ اللَّهِ حُكْمًا لِّقَوْمٍ يُوقِنُونَ> \\
& Is it the judgment of pre-Islamic ignorance they seek? Who could be a better judge than Allah for people of sure faith? \\
24 & \<وَيَقُولُونَ مَتَىٰ هَٰذَا الْوَعْدُ إِن كُنتُمْ صَادِقِينَ> \\
& And they ask the believers, “When will this threat come to pass, if what you say is true?” \\
\bottomrule
\end{tabular}
\caption{The top shared Quran verses in the data collection of Iran along with their translation from \url{quran.com}. For some long verses, only the fragments commonly shared in tweets are added.}
\label{tab:top_quran_iran}
\end{table*}

\begin{table*}
\centering
\begin{tabular}{ p{1cm} R{16cm}}
\toprule
\multicolumn{1}{l}{\textbf{Count}} & \textbf{Top Quran Verses/Fragments (Coordination Datasets)}\hspace{0cm}\mbox{ } \\ 
\midrule
5129 & \< وَيُطْعِمُونَ الطَّعَامَ عَلَى حُبِّهِ مِسْكِينًا وَيَتِيمًا وَأَسِيرًا> \\
& and give food despite their desire for it to the poor, the orphan, and the captive. \\
4125 & \<قُلْ مَا أَنفَقْتُم مِّنْ خَيْرٍ فَلِلْوَالِدَيْنِ وَالأَقْرَبِينَ وَالْيَتَامَى وَالْمَسَاكِينِ وَابْنِ السَّبِيلِ> \\
& Whatever donations you give are for parents, relatives, orphans, the poor, and needy travelers. \\
\bottomrule
\end{tabular}
\caption{The top shared Quran verses in the data collection of coordinated propaganda along with their translation from \url{quran.com}. For some long verses, only the fragments commonly shared in tweets are added.}
\label{tab:top_quran_coordinated}
\end{table*}

\FloatBarrier

\section{Top shared Hadith}
\label{appendixHadith}

\begin{table}[H]
\centering
\begin{tabular}{ p{1cm} R{7cm}}
\toprule
\multicolumn{1}{l}{\textbf{Count}} & \textbf{Top Authentic Hadith (KSA)}\hspace{0cm}\mbox{ } \\ 
\midrule
3113 & \hspace{0.9cm}\<لَأنْ أقولَ : سُبحانَ اللهِ والحمدُ للهِ ولا إلهَ إلَّا اللهُ \\ واللهُ أكبَرُ أحَبُّ إليَّ ممَّا طلَعَتْ عليه الشَّمسُ> \\
& For me to say: (Glory is to Allah,and praise is to Allah,and there is none worthy of worship but Allah,and Allah is the Most Great) is dearer to me than all that the sun rises upon (i.e. the whole world). \\
2114 & \hspace{0.3cm}\<كَلِمَتَانِ خَفِيفَتَانِ علَى اللِّسَانِ، ثَقِيلَتَانِ في المِيزَانِ، حَبِيبَتَانِ \\ إلى الرَّحْمَنِ: سُبْحَانَ اللهِ  وَبِحَمْدِهِ، سُبْحَانَ اللهِ العَظِيمِ.> \\
&  Two words are light on the tongue, weigh heavily in the balance, and are loved by the Most Merciful One: Glorified is Allah and praised is He, Glorified is Allah the Most Great. \\
1775 & \hspace{0.4cm}\<ما مِنْ عبدٍ يسترْعيه اللهُ رعيَّةً، يموتُ يومَ يموتُ ، وهوَ \\ غاشٌّ لرعِيَّتِهِ، إلَّا حرّمَ اللهُ عليْهِ الجنَّةَ> \\
&  Any slave whom Allah makes him in charge of subjects and he dies while he is not sincere to them, Allah will make Jannah unlawful for him \\
\bottomrule
\end{tabular}
\caption{The top shared authentic hadiths in the data collection of Iran along with their translation from \url{sunnah.com}.}
\label{tab:top_hadith_a_ksa}
\end{table}

\begin{table}[H]
\centering
\begin{tabular}{ p{1cm} R{7cm}}
\toprule
\multicolumn{1}{l}{\textbf{Count}} & \textbf{Top Good/Weak Hadith (KSA)}\hspace{0cm}\mbox{ } \\ 
\midrule
1090 & \<يا حازمُ أَكْثِر من قولِ لا حولَ ولا قوَّةَ إلَّا باللَّهِ فإنَّها من \\ كُنوزِ الجنَّةِ> \\
& O Hazim, say often: there is no power and no strength except with Allah, for it is one of the treasures of Paradise. \\
\bottomrule
\end{tabular}
\caption{The top shared good and weak hadiths in the data collection of KSA along with their translation from \url{sunnah.com}.}
\label{tab:top_hadith_l_ksa}
\end{table}

\begin{table}[H]
\centering
\begin{tabular}{ p{1cm} R{7cm}}
\toprule
\multicolumn{1}{l}{\textbf{Count}} & \textbf{Top fabricated Hadith (KSA)}\hspace{0cm}\mbox{ } \\ 
\midrule
355 & \<من قرأ سورة الدخان في ليلة غفر له ما تقدم من ذنبه> \\
& Whoever recites Surah Ad-Dukhan during the night, his previous sins will be forgiven.\\
\bottomrule
\end{tabular}
\caption{The top shared fabricated hadiths in the data collection of KSA along with their translation.}
\label{tab:top_hadith_f_ksa}
\end{table}

\begin{table*}
\centering
\begin{tabular}{ p{1cm} R{16cm}}
\toprule
\multicolumn{1}{l}{\textbf{Count}} & \textbf{Top Authentic Hadith (UAE)}\hspace{0cm}\mbox{ } \\ 
\midrule
281 & \hspace{0.8cm}\<لَأنْ أقولَ : سُبحانَ اللهِ والحمدُ للهِ ولا إلهَ إلَّا اللهُ واللهُ أكبَرُ أحَبُّ إليَّ ممَّا طلَعَتْ عليه الشَّمسُ> \\
& For me to say: (Glory is to Allah,and praise is to Allah,and there is none worthy of worship but Allah,and Allah is the Most Great) is dearer to me than all that the sun rises upon (i.e. the whole world). \\
176 & \hspace{0.5cm}\<ينزلُ اللهُ كلَّ ليلةٍ إلى السماءِ الدنيا ، حين يبقى ثلثُ الليلِ الآخرِ ،فيقولُ: من يدعوني فأستجيبُ له ، من يسألني فأعطيه ، من \\ يستغفرُني  فأغفرُ له> \\
&  Our Lord, the Blessed and the Exalted, descends every night to the lowest heaven when one-third of the latter part of the night is left, and says: Who supplicates Me so that I may answer him? Who asks Me so that I may give to him? Who asks Me forgiveness so that I may forgive him? \\
170 & \hspace{1.1cm}\<كَلِمَتَانِ خَفِيفَتَانِ علَى اللِّسَانِ، ثَقِيلَتَانِ في المِيزَانِ، حَبِيبَتَانِ إلى الرَّحْمَنِ: سُبْحَانَ اللهِ وَبِحَمْدِهِ، سُبْحَانَ اللهِ العَظِيمِ.> \\
&  Two words are light on the tongue, weigh heavily in the balance, and are loved by the Most Merciful One: Glorified is Allah and praised is He, Glorified is Allah the Most Great. \\
\bottomrule
\end{tabular}
\caption{The top shared authentic hadiths in the data collection of UAE along with their translation from \url{sunnah.com}.}
\label{tab:top_hadith_a_uae}
\end{table*}

\begin{table*}
\centering
\begin{tabular}{ p{1cm} R{16cm}}
\toprule
\multicolumn{1}{l}{\textbf{Count}} & \textbf{Top Good/Weak Hadith (UAE)}\hspace{0cm}\mbox{ } \\ 
\midrule
87 & \hspace{0.5cm} \<خيرُ الدعاءِ دعاءُ يومِ عرفةَ، وخيرُ ما قلتُ أنا والنبيُّونَ من قبلي: لا إلهَ إلَّا اللهُ وحدَهُ لا شريكَ له، لهُ الملكُ ولهُ الحمدُ، وَهوَ على \\ كلِّ شيءٍ قديرٌ> \\
& The best invocation is that of the Day of Arafah, and the best that anyone can say is what I and the Prophets before me have said: None has the right to be worshipped but Allah Alone, Who has no partner. His is the dominion and His is the praise, and He is Able to do all things. \\
78 & \<يا حازمُ أَكْثِر من قولِ لا حولَ ولا قوَّةَ إلَّا باللَّهِ فإنَّها من كُنوزِ الجنَّةِ> \\
& O Hazim, say often: there is no power and no strength except with Allah, for it is one of the treasures of Paradise. \\
\bottomrule
\end{tabular}
\caption{The top shared good and weak hadiths in the data collection of UAE along with their translation from \url{sunnah.com}.}
\label{tab:top_hadith_l_uae}
\end{table*}

\begin{table}
\centering
\begin{tabular}{ p{1cm} R{7cm}}
\toprule
\multicolumn{1}{l}{\textbf{Count}} & \textbf{Top fabricated Hadith (UAE)}\hspace{0cm}\mbox{ } \\ 
\midrule
26 & \<من قرأ سورة الدخان في ليلة غفر له ما تقدم من ذنبه> \\
& Whoever recites Surah Ad-Dukhan during the night, his previous sins will be forgiven.\\
2 & \<أبغض الكلام إلى الله الفارسية> \\
& The Persian language is most hated talk to God.\\
\bottomrule
\end{tabular}
\caption{The top shared fabricated hadiths in the data collection of UAE along with their translation.}
\label{tab:top_hadith_f_uae}
\end{table}


\begin{table}[H]
\centering
\begin{tabular}{ p{1cm} R{7cm}}
\toprule
\multicolumn{1}{l}{\textbf{Count}} & \textbf{Top Authentic Hadith (Egypt)}\hspace{0cm}\mbox{ } \\ 
\midrule
455 & \<إنَّ أشدَّ النَّاسِ عذابًا عندَ اللَّهِ المصَوِّرونَ> \\
& Among the people who will be most severely punished on the Day of Resurrection will be the image-makers. \\
355 & \hspace{0.8cm}\<لَأنْ أقولَ : سُبحانَ اللهِ والحمدُ للهِ ولا إلهَ إلَّا اللهُ \\ واللهُ أكبَرُ أحَبُّ إليَّ ممَّا طلَعَتْ عليه الشَّمسُ> \\
& For me to say: (Glory is to Allah,and praise is to Allah,and there is none worthy of worship but Allah,and Allah is the Most Great) is dearer to me than all that the sun rises upon (i.e. the whole world). \\
\bottomrule
\end{tabular}
\caption{The top shared authentic hadiths in the data collection of Egypt along with their translation from \url{sunnah.com}.}
\label{tab:top_hadith_a_egypt}
\end{table}

\begin{table}[H]
\centering
\begin{tabular}{ p{1cm} R{7cm}}
\toprule
\multicolumn{1}{l}{\textbf{Count}} & \textbf{Top Good/Weak Hadith (Egypt)}\hspace{0cm}\mbox{ } \\ 
\midrule
253 & \<إِنَّ اللَّهَ يَبْغَضُ البَلِيغَ مِنَ الرِّجَالِ الَّذِي يَتَخَلَّلُ بِلِسَانِهِ كَمَا  \\ تَتَخَلَّلُ البَقَرَةُ> \\
& Allah dislikes an eloquent person who rolls his tongue as a cow rolls its tongue.\\
249 & \< لا يُبلغُني أحدٌ من أصحابي عن أحدٍ شيئًا؛ فإني أحب أن \\ أخرج إليكم وأنا سليمُ الصدرِ> \\
& None of my companions should tell me anything about anyone, for I would like to come out to you with a sound heart. \\
240 & \hspace{0.6cm} \< إذا أصبح ابنُ آدمَ ؛ فإنَّ الأعضاءَ كلَّها تكفِّر اللسانَ، \\ فتقول : اتقِ اللهَ فينا ؛  فإنما نحن بك ؛فإن استقمت \hspace{0.7cm}\\ استقمْنا وإنِ اعوججتَ اعوججنا> \\
& When the son of Adam gets up in the morning, all the limbs humble themselves before the tongue and say: 'Fear Allah for our sake because we are with you: if you are straight, we will be straight; and if you are crooked, we will become crooked.' \\
236 & \<رضا الرَّبِّ في رضا الوالدِ وسَخَطُ الرَّبِّ في سَخَطِ الوالدِ> \\
& 
The pleasure of God is in the pleasure of the parent, and the wrath of God is in the wrath of the parent. \\
\bottomrule
\end{tabular}
\caption{The top shared good and weak hadiths in the data collection of Egypt along with their translation.}
\label{tab:top_hadith_gw_egypt}
\end{table}

\begin{table}[H]
\centering
\begin{tabular}{ p{1cm} R{7cm}}
\toprule
\multicolumn{1}{l}{\textbf{Count}} & \textbf{Top fabricated Hadith (Egypt)}\hspace{0cm}\mbox{ } \\ 
\midrule
96 & \<من قرأ سورة الدخان في ليلة غفر له ما تقدم من ذنبه> \\
& Whoever recites Surah Ad-Dukhan during the night, his previous sins will be forgiven.\\
10 & \<أبغض الكلام إلى الله الفارسية> \\
& The Persian language is most hated talk to God.\\
\bottomrule
\end{tabular}
\caption{The top shared fabricated hadiths in the data collection of Egypt along with their translation.}
\label{tab:top_hadith_f_egypt}
\end{table}

\begin{table}[H]
\centering
\begin{tabular}{ p{1cm} R{7cm}}
\toprule
\multicolumn{1}{l}{\textbf{Count}} & \textbf{Top Authentic Hadith (Iran)}\hspace{0cm}\mbox{ } \\ 
\midrule
30 & \<المُؤْمِنُ لِلْمُؤْمِنِ كَالْبُنْيَانِ يَشُدُّ بعضُه بَعْضًا> \\
& A faithful believer to a faithful believer is like the bricks of a wall, reinforcing each other. \\
19 & \<مَنْ صَلَّى عَلَيَّ صَلَاةً، صَلَّى اللهُ عَلَيْهِ بِهَا عَشْرًا> \\
& Whoever prays for Allah's blessings upon me once, will be blessed for it by Allah ten times. \\
18 & \<من سَرَّه أن ينظرَ إلى رجلٍ من أهلِ الجنةِ فلينظرْ إلى هذا> \\
& He who wants to see a man from the dwellers of Jannah, let him look at him. \\
17 & \<مَن صلى عَلَيَّ واحدةً ، صلى اللهُ عليه بها عَشْرًا> \\
& Whoever prays for Allah's blessings upon me once, will be blessed for it by Allah ten times \\
13 & \hspace{0.5cm} \<مثلُ المؤمنين في تَوادِّهم ، وتَرَاحُمِهِم ، وتعاطُفِهِمْ. مثلُ \\ الجسَدِ إذا اشتكَى منْهُ  عضوٌ تدَاعَى لَهُ سائِرُ الجسَدِ بالسَّهَرِ \hspace{0.2cm}\\ والْحُمَّى> \\
& The believers in their mutual kindness, compassion and sympathy are just like one body. When one of the limbs suffers, the whole body responds to it with wakefulness and fever \\
\bottomrule
\end{tabular}
\caption{The top shared authentic hadiths in the data collection of Iran along with their translation from \url{sunnah.com}.}
\label{tab:top_hadith_a_iran}
\end{table}

\begin{table}[H]
\centering
\begin{tabular}{ p{1cm} R{7cm}}
\toprule
\multicolumn{1}{l}{\textbf{Count}} & \textbf{Top Good/Weak Hadith (Iran)}\hspace{0cm}\mbox{ } \\ 
\midrule
17 & \hspace{0.2cm} \<حُسَيْنٌ مِنِّي وَأَنَا مِنْ حُسَيْنٍ. أَحَبَّ اللَّهُ مَنْ أَحَبَّ حُسَيْنًا.\\ حُسَيْنٌ سِبْطٌ مِنَ الأَسْبَاطِ> \\
& Husain is from me, and I am from Husain. Allah loves whoever loves Husain. Husain's offspring will become a great tribe. \\
15 & \<البخيلُ الذي من ذُكِرْتُ عندَه فلم يُصَلِّ عليَّ> \\
& The miser is the one in whose presence I am mentioned but he does not supplicate for me. \\
15 & \<الحسَنُ والحُسَيْنُ سيِّدا شبابِ أهلِ الجنَّةِ> \\
& Al-Hasan and Al-Husain are the chiefs of the youths of Paradise. \\
\bottomrule
\end{tabular}
\caption{The top shared good and weak hadiths in the data collection of Iran along with their translation from \url{sunnah.com}.}
\label{tab:top_hadith_l_iran}
\end{table}

\begin{table}[H]
\centering
\begin{tabular}{ p{1cm} R{7cm}}
\toprule
\multicolumn{1}{l}{\textbf{Count}} & \textbf{Top fabricated Hadith (Iran)}\hspace{0cm}\mbox{ } \\ 
\midrule
77 & \<هي لك يا علي> \\
& She is yours, Ali\\
36 & \hspace{0.25cm} \<أمرنا رسول الله صلى الله عليه وسلم أن نعرض أولادنا \\ على حب علي بن أبى طالب> \\
& The Messenger of Allah ordered us to present our children to the love of Ali ibn Abi Talib \\
36 & \hspace{0.15cm} \<من أزهد الناس في العالم فقال صلى الله عليه وآله وسلم \\ أهل بيته>  \\
& When asked who are the most ascetic person in the world? The prophet referred to the people of his house.\\
\bottomrule
\end{tabular}
\caption{The top shared fabricated hadiths in the data collection of Iran along with their translation.}
\label{tab:top_hadith_f_iran}
\end{table}


\begin{table*}
\centering
\begin{tabular}{ p{1cm} R{16cm}}
\toprule
\multicolumn{1}{l}{\textbf{Count}} & \textbf{Top Authentic Hadith (Coordination Datasets)}\hspace{0cm}\mbox{ } \\ 
\midrule
863 & \hspace{1.1cm}\<كَلِمَتَانِ خَفِيفَتَانِ علَى اللِّسَانِ، ثَقِيلَتَانِ في المِيزَانِ، حَبِيبَتَانِ إلى الرَّحْمَنِ: سُبْحَانَ اللهِ وَبِحَمْدِهِ، سُبْحَانَ اللهِ العَظِيمِ.> \\
&  Two words are light on the tongue, weigh heavily in the balance, and are loved by the Most Merciful One: Glorified is Allah and praised is He, Glorified is Allah the Most Great. \\
\bottomrule
\end{tabular}
\caption{The top shared authentic hadiths in the data collection of coordinated propaganda along with their translation.}
\label{tab:top_hadith_a_coordinated}
\end{table*}

\begin{table*}
\centering
\begin{tabular}{ p{1cm} R{16cm}}
\toprule
\multicolumn{1}{l}{\textbf{Count}} & \textbf{Top Good/Weak Hadith (Coordination Datasets)}\hspace{0cm}\mbox{ } \\ 
\midrule
87 & \hspace{0.5cm} \<خيرُ الدعاءِ دعاءُ يومِ عرفةَ، وخيرُ ما قلتُ أنا والنبيُّونَ من قبلي: لا إلهَ إلَّا اللهُ وحدَهُ  لا شريكَ له، لهُ الملكُ ولهُ الحمدُ، وَهوَ على \\ كلِّ شيءٍ قديرٌ> \\
& The best invocation is that of the Day of Arafah, and the best that anyone can say is what I and the Prophets before me have said: None has the right to be worshipped but Allah Alone, Who has no partner. His is the dominion and His is the praise, and He is Able to do all things. \\
\bottomrule
\end{tabular}
\caption{The top shared good/weak hadiths in the data collection of coordinated propaganda along with their translation.}
\label{tab:top_hadith_gw_coordinated}
\end{table*}

\begin{table*}
\centering
\begin{tabular}{ p{1cm} R{16cm}}
\toprule
\multicolumn{1}{l}{\textbf{Count}} & \textbf{Top fabricated Hadith (Coordination Datasets)}\hspace{0cm}\mbox{ } \\ 
\midrule
243 & \<من قرأ سورة الدخان في ليلة غفر له ما تقدم من ذنبه> \\
& Whoever recites Surah Ad-Dukhan during the night, his previous sins will be forgiven.\\
50 & \<إنما الباذنجان شفاء لكل داء> \\
& Eggplant is a cure for every disease.\\
\bottomrule
\end{tabular}
\caption{The top shared fabricated hadiths in the data collection of coordinated propaganda along with their translation.}
\label{tab:top_hadith_f_coordinated}
\end{table*}

\newpage

\section{Political Tweets in Original Language}
\label{sec:political-arabic}

\begin{figure}[H]
    \centering
    \begin{subfigure}[b]{0.49\linewidth}
        \centering
        \includegraphics[width=\linewidth]{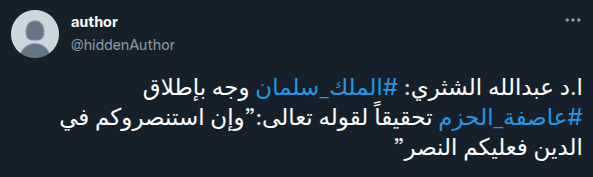}
        \caption{politicized Quran (KSA)}
        \label{fig:ksa-quran1}
    \end{subfigure}
    \hfill
    \begin{subfigure}[b]{0.49\linewidth}
        \centering
        \includegraphics[width=\linewidth]{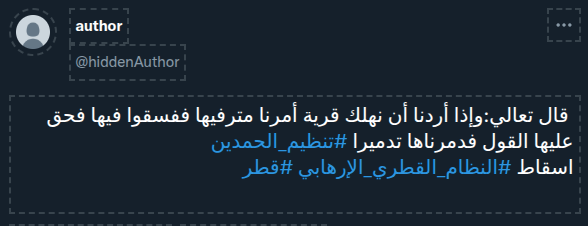}
        \caption{politicized Quran (KSA)}
        \label{fig:ksa-quran2}
    \end{subfigure}
    \caption{Examples of politicized Quran in the Saudi data collection}
    \label{fig:combined-ksa-q}
\end{figure}

\begin{figure}[H] 
    \centering
    \begin{subfigure}[b]{0.49\linewidth}
        \centering
        \includegraphics[width=\linewidth]{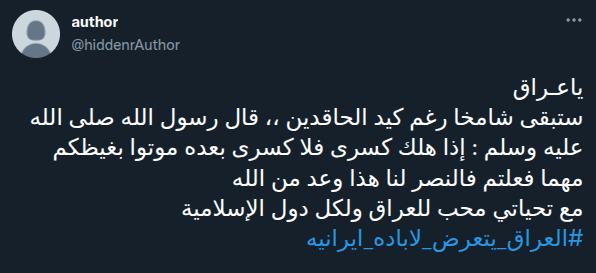}
        \caption{politicized hadith (KSA)}
        \label{fig:ksa-hadith1}
    \end{subfigure}
    \hfill
    \begin{subfigure}[b]{0.49\linewidth}
        \centering
        \includegraphics[width=\linewidth]{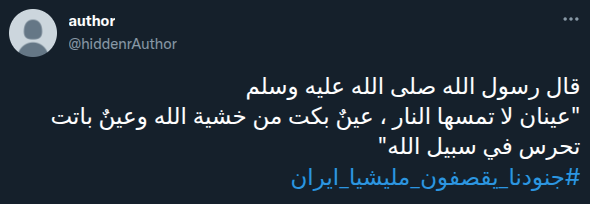}
        \caption{politicized hadith (KSA)}
        \label{fig:ksa-hadith2}
    \end{subfigure}
    \caption{Examples of politicized hadith in the Saudi data collection.}
    \label{fig:combined-ksa-h}
\end{figure}

\begin{figure}[H]
    \centering
    \begin{subfigure}[b]{0.49\linewidth}
        \centering
        \includegraphics[width=\linewidth]{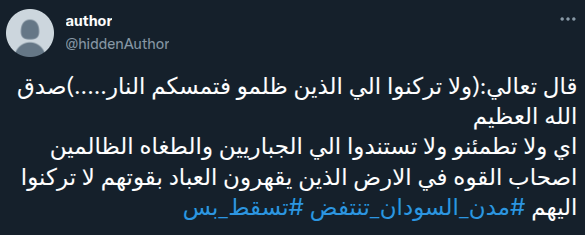}
        \caption{politicized Quran (UAE)}
        \label{fig:uae-quran1}
    \end{subfigure}
    \hfill
    \begin{subfigure}[b]{0.49\linewidth}
        \centering
        \includegraphics[width=\linewidth]{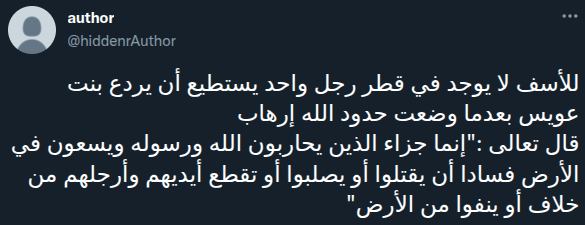}
        \caption{politicized Quran (UAE)}
        \label{fig:uae-quran2}
    \end{subfigure}
    \caption{Examples of politicized Quran in the Emirati data collection.}
    \label{fig:combined-uae-q}
\end{figure}

\begin{figure}[H]
    \centering
    \begin{subfigure}[b]{0.45\linewidth}
        \centering
        \includegraphics[width=\linewidth]{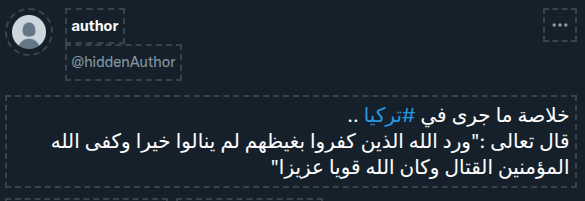}
        \caption{politicized Quran (Egypt)}
        \label{fig:egypt-quran1}
    \end{subfigure}
    \hfill
    \begin{subfigure}[b]{0.45\linewidth}
        \centering
        \includegraphics[width=\linewidth]{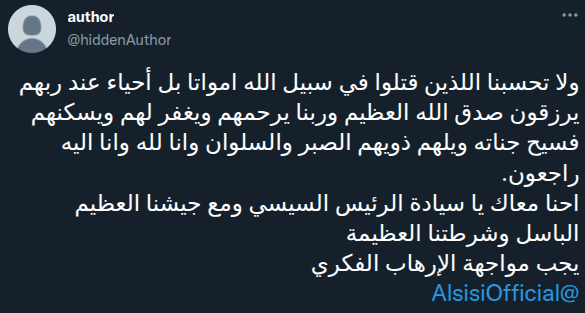}
        \caption{politicized Quran (Egypt)}
        \label{fig:egypt-quran2}
    \end{subfigure}
    \caption{Examples of politicized Quran in the Egyptian data collection.}
    \label{fig:combined-egypt-q}
\end{figure}

\begin{figure}[H]
    \centering
    \begin{subfigure}[b]{0.45\linewidth}
        \centering
        \includegraphics[width=\linewidth]{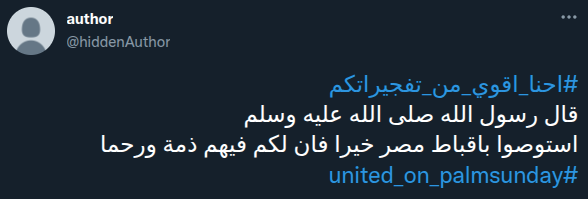}
        \caption{politicized hadith (Egypt)}
        \label{fig:egypt-hadith1}
    \end{subfigure}
    \hfill
    \begin{subfigure}[b]{0.45\linewidth}
        \centering
        \includegraphics[width=\linewidth]{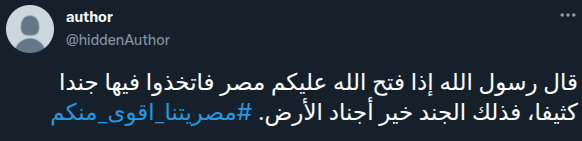}
        \caption{politicized hadith (Egypt)}
        \label{fig:egypt-hadith2}
    \end{subfigure}
    \caption{Examples of politicized hadith in the Egyptian data collection.}
    \label{fig:combined-egypt-h}
\end{figure}

\begin{figure}[H]
    \centering
    \begin{subfigure}[b]{0.45\linewidth}
        \centering
        \includegraphics[width=\linewidth]{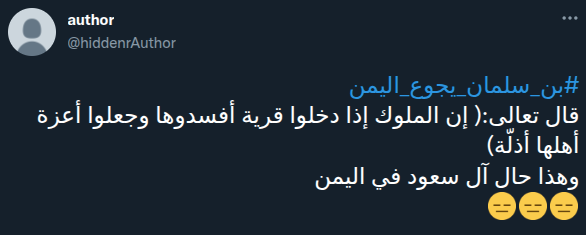}
        \caption{politicized Quran (Iran)}
        \label{fig:iran-quran1}
    \end{subfigure}
    \hfill
    \begin{subfigure}[b]{0.45\linewidth}
        \centering
        \includegraphics[width=\linewidth]{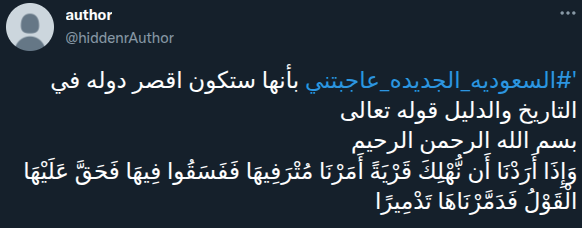}
        \caption{politicized Quran (Iran)}
        \label{fig:iran-quran2}
    \end{subfigure}
    \caption{Examples of politicized hadith in the Iranian data collection.}
    \label{fig:combined-iran-q}
\end{figure}

\begin{figure}[H]
    \centering
    \begin{subfigure}[b]{0.45\linewidth}
        \centering
        \includegraphics[width=\linewidth]{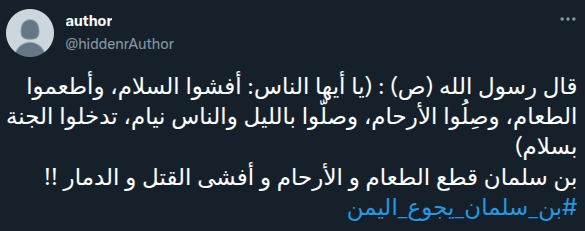}
        \caption{politicized hadith (Iran)}
        \label{fig:iran-hadith1}
    \end{subfigure}
    \hfill
    \begin{subfigure}[b]{0.45\linewidth}
        \centering
        \includegraphics[width=\linewidth]{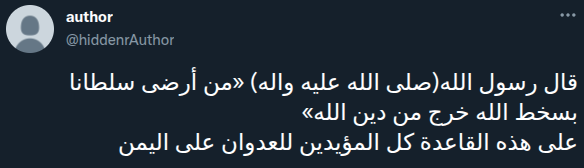}
        \caption{politicized hadith (Iran)}
        \label{fig:iran-hadith2}
    \end{subfigure}
    \caption{Examples of politicized hadith in the Iranian data collection.}
    \label{fig:combined-iran-h}
\end{figure}

\begin{figure}[H]
    \centering
    \begin{subfigure}[b]{0.45\linewidth}
        \centering
        \includegraphics[width=\linewidth]{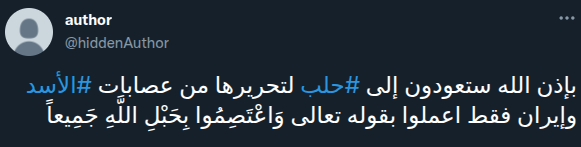}
        \caption{politicized Quran (Coordination Datasets)}
        \label{fig:coordination-quran}
    \end{subfigure}
    \hfill
    \begin{subfigure}[b]{0.45\linewidth}
        \centering
        \includegraphics[width=\linewidth]{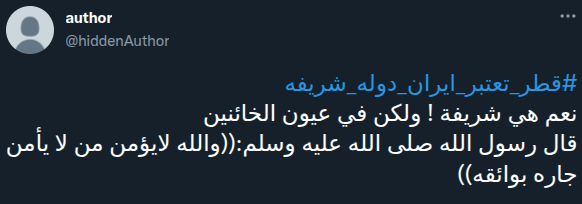}
        \caption{politicized hadith (Coordination Datasets)}
        \label{fig:coordination-hadith}
    \end{subfigure}
    \caption{Examples of politicized Quran and hadith in the coordination data collection.}
    \label{fig:combined-coordination}
\end{figure}

\begin{figure}[H]
    \centering
    \begin{subfigure}[b]{0.45\linewidth}
        \centering
        \includegraphics[width=\linewidth]{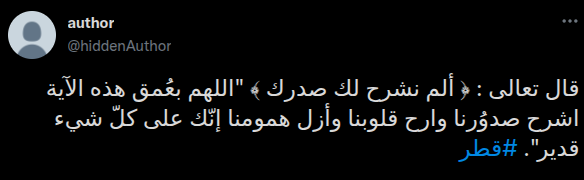}
        \caption{non-political Quran (Coordination Datasets)}
        \label{fig:coordination-quran}
    \end{subfigure}
    \hfill
    \begin{subfigure}[b]{0.45\linewidth}
        \centering
        \includegraphics[width=\linewidth]{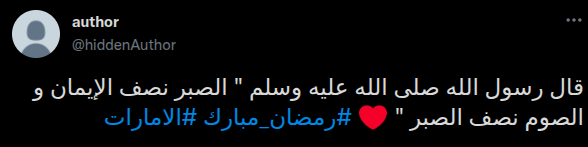}
        \caption{non-political hadith (Coordination Datasets)}
        \label{fig:coordination-hadith}
    \end{subfigure}
    \caption{Examples of non-political Quran and hadith in the coordination data collection.}
    \label{fig:combined-coordination}
\end{figure}

\end{document}